\documentclass[a4paper,fleqn,usenatbib]{mnras}

\pdfoutput=1

\usepackage{txfonts}

\usepackage[T1]{fontenc}
\usepackage{ae,aecompl}

\usepackage{graphicx}	% Including figure files
\usepackage{amssymb}	% Extra maths symbols
\usepackage[flushleft]{threeparttable} % for table 1, allows table footnotes underneath the table to explain what each parameter is
\usepackage{float}

\newcommand{\rpzvt}{\left (R,\,\phi, \,z,\,v_z,\,t\right )}

\newcommand{\hy}{H\,{\sc i}~}

\title[Bending waves in Milky Way-like discs]{Spontaneous generation of bending waves in isolated Milky Way-like discs}

\author[Matthew H. Chequers and Lawrence M. Widrow]
{Matthew H. Chequers\thanks{E-mail: 12mhc@queensu.ca} and Lawrence M. Widrow\thanks{E-mail: widrow@queensu.ca}
\\
Department of Physics, Engineering Physics \& Astronomy, Queen's
University, Kingston, ON K7L 3N6, Canada}

\date{Accepted XXX. Received YYY; in original form ZZZ}

\pubyear{2017}

\begin{document}
\def\aligned{\vcenter\bgroup\let\\\cr
\halign\bgroup&\hfil${}##{}$&${}##{}$\hfil\cr}
\def\endaligned{\crcr\egroup\egroup}

\label{firstpage}
\pagerange{\pageref{firstpage}--\pageref{lastpage}}
\maketitle

% Abstract of the paper

\begin{abstract}

\noindent
We study the spontaneous generation and evolution of bending waves in $N$-body
simulations of two isolated Milky Way-like galaxy models. The models differ by their disc-to-halo mass ratios, and hence by their susceptibility to the formation of a bar and spiral structure. Seeded from shot noise
in the particle distribution, bending waves rapidly form in both models and
persist for many billions of years. Waves at intermediate radii manifest as corrugated structures in vertical position and velocity that are
tightly wound, morphologically leading, and dominated by the $m=1$
azimuthal Fourier component.  A spectral analysis of the waves suggests they
are a superposition of modes from two continuous branches in the
Galactocentric radius-rotational frequency plane.  The lower-frequency branch
is dominant and is responsible for the corrugated, leading, and warped
structure. Over time, power in this branch migrates outward,
lending credence to an inside-out formation scenario for the warp.  Our power
spectra qualitatively agree with results from linear perturbation theory and
a WKB analysis, both of which include self-gravity. Thus, we conclude
that the waves in our simulations are self-gravitating and not purely
kinematic. These waves are reminiscent of the wave-like pattern recently found in Galactic star counts from
the Sloan Digital Sky Survey and smoothly transition to a warp near
the disc's edge. Velocity measurements from \textit{Gaia} data will be instrumental in testing the true wave nature of the corrugations.  We also compile a list of ``minimum requirements" needed to observe bending waves in external galaxies. 

\end{abstract}

% Select between one and six entries from the list of approved keywords.
% Don't make up new ones.


%%%%%%%%%%%%%%%%%%%%%%%%%%%%%%%%%%%%%%%%%%%%%%%%%%

%%%%%%%%%%%%%%%%% BODY OF PAPER %%%%%%%%%%%%%%%%%%

\section{Introduction}

It is common for disc galaxies to bend in and out of their respective
midplanes.  The most conspicuous examples are the warps seen in the
outer regions of both \hy and stellar discs (see \citealt{binney1992} and \citealt{sellwood2013}, and references therein).  There are classic ``integral sign" warps as
well as warps with more complicated morphologies.  Indeed, detailed
H\,{\sc i} maps of the Milky Way reveal a complex pattern of warping
and flaring beyond the Solar circle \citep{levine2006}. More recently, \citet{reyle2009} showed that warped distortions similar to that in the gas also exist in the stellar \textit{and} dust components of the Milky Way (see also \citealt{cox1996} for an earlier discussion of warps in the stellar and gaseous components of external discs).

Warps are but one example of how a galactic disc might deviate from
planarity.  For example, a disc may also exhibit corrugations or
wave-like structures in the direction normal to its midplane.  Such
structures are most easily observed in the Milky Way where we have
access to the full six-dimensional phase space.  Consider the
Monoceros ring, an overdensity in stars that arcs around the Galactic
centre at a Galactocentric radius of about 18 kpc 
\citep{newberg2002,yanny2003,li2012,morganson2016}.  The Monoceros ring was
originally thought to comprise tidal debris from a disrupted dwarf
galaxy \citep{martin2004,penarrubia2005}.  The alternative is that it
is a feature intrinsic to the disc.  For example, several authors have
pointed out that the Monoceros ring arose from warping and
flaring of the disc \citep{momany2004,momany2006, hammersley2011} or resulted from the disc interacting with a massive satellite
\citep{kazantzidis2008,younger2008,purcell2011,gomez2016}. More
recently, \citet{xu2015} suggested that the Monoceros ring is one of
several crests of an oscillatory bending wave observed in the
direction of the Galactic anti-centre.

The counterpart to these Galactic waves may have been observed in
external galaxies.  For example, corrugations in the line-of-sight
velocity field have been detected in the H$\alpha$ emission of nearly
face-on spiral galaxies by \citet{alfaro2001} and
\citet{sanchez-gil2015}.  These wave-like variations in the vertical
velocity field appear to coincide with large amplitude spiral arms.
One explanation is that they arise from the interaction between spiral
density waves and the galaxy's gaseous disc via the hydraulic bore
mechanism \citep{martos1998,martos1999}.  On the other hand,
simulations show that corrugated velocity patterns in the vertical
velocity field arise in both stellar and gaseous discs
\citep{gomez2016,gomez2017}, a result that suggests a gravitational
rather than hydrodynamic origin.

Departures from planarity in the disc of the Milky Way often appear as
asymmetries in the phase space distribution of stars North and South
of the Galactic midplane.  For example, the discovery of bending waves
by \citet{xu2015} involved mapping the North-South number count
asymmetry as a function of position in the disc plane.  Evidence for a
North-South asymmetry in the vertical profile of number counts for a
$1\,{\rm kpc}$ cylinder centered on the Sun had already been found by
\citet{widrow2012} and \citet{yanny2013} (see \citealt{ferguson2017} for a more recent confirmation using a larger photometric sample). Furthermore,
\citet{widrow2012}, \citet{williams2013}, \citet{carlin2013}, and
\citet{sun2015} observed bulk motions in Solar Neighbourhood disc stars perpendicular to
the midplane (vertical motions).  These motions can be interpreted as a
combination of the bending and breathing modes that are theoretically predicted for a plane-symmetric system \citep{mathur1990,
  weinberg1991, gomez2013, widrow2014,widrow2015,gomez2016,gomez2017}.

The existence of warps has been attributed to the tidal effects of
satellite galaxies, interactions of the disc with its dark halo,
internal excitation of bending instabilities in gaseous discs,
and intergalactic winds and magnetic fields, to name just a few
\citep{kahn1959, hunter1969, sparke1984, sparke1988, battaner1990,
  binney1992, debattista1999, lopez2002, revaz2004, shen2006}.
Satellites and halo substructure have been invoked to explain the
bending and breathing waves seen in the Milky Way
\citep{widrow2012,gomez2013,widrow2014,feldmann2015,d'onghia2016,gomez2016,gomez2017},
though these features can also be generated by spiral structure
\citep{debattista2014,faure2014,monari2016a}, the bar
\citep{monari2015}, or some (non-linear) combination of the two
\citep{monari2016b}.

The conclusion is that a galactic disc continually experiences
perturbations that can set up waves perpendicular to its midplane.
The question remains as to what happens to these waves once they are
produced.  Are they long-lived or do they quickly decohere, heating
and thickening the disc?  Can the bending waves be understood as purely
kinematic structure, as has been suggested by \citet{delavega2015}, or
is self-gravity essential to their evolution?
Finally, is there a connection between the corrugations of
\citet{xu2015} and the bulk motions in the Solar neighbourhood on the
one hand, and the warp at the edge of the Galactic disc on the other?

In this paper, we address these and other questions through a series
of idealized isolated galaxy simulations of Milky Way-like models, as well as theoretical
arguments based on both the eigenmode and WKB analysis of bending
waves by \citet{hunter1969}, \citet{sparke1984}, \citet{sparke1988}, and \citet{nelson1995}. The bending waves in our simulations arise without any provocation apart from the random noise of the particle distributions used to characterize the (smooth) halo, bulge, and disc. We emphasize that our results are specific to Milky Way-like galaxies and may not generalize to other stellar discs. Indeed, there are very thin disc galaxies that show no sign of bending. These galaxies, which may well be low surface brightness galaxies viewed edge-on \citep{bizyaev2017}, may not be as susceptible to the dynamical mechanisms described in this work as the galaxies considered here.

Attempts to understand the evolution of galactic bending waves in
general, and warps in particular, date back over half a century.  Much
of the discussion parallels efforts to understand spiral structure.
In particular, it was recognized early on that in the absence of
self-gravity, an initial warp will shear due to differential
precession \citep{kahn1959} in a process akin to the winding problem
for kinematic spiral structure.  However, \citet{lyndenbell1965}
argued that a self-gravitating disc might support true bending modes,
which makes warps long-lived.  This idea was pursued in
detail by \citet{hunter1969} who found that in general, an isolated
disc supports a continuum of modes and therefore a generic bending
perturbation will disperse.  For a disc embedded in
a spherically symmetric halo, the only discrete ``mode" is the trivial zero-frequency tilt of the disc as a whole.  If, on the
other hand, the disc is embedded in a flattened halo, then the
tilt mode is distorted but remains discrete and is therefore a
candidate for explaining long-lived warps.  This idea, first proposed
by \citet{dekel1983} and \citet{toomre1983}, was studied in detail by
\citet{sparke1984} and \citet{sparke1988} who treated the disc as a
system of concentric rings embedded in the static potential of a
flattened halo (see \citealt{revaz2001} for a physical justification of this ring model). 

One of the main limitations in \citet{sparke1984} and
\citet{sparke1988} is that the halo is treated as a static potential,
whereas live haloes can respond to the time-evolving gravitational
field of the disc.  In addition, the disc will set up wakes in the
halo, which then act back on the disc via dynamical friction
\citep{chandra1943}.  \citet{bertin1980} argued that the halo-disc
interaction can actually excite bending waves that are inside their
co-rotation radius, though waves outside co-rotation are damped.  Their
analysis made a number of approximations, most notably that the halo
comprised a spatially uniform Maxwellian distribution of
particles. Also, they did not consider the halo contribution to the
vertical restoring force.  \citet{nelson1995} re-examined these claims
in the context of more realistic halo models and concluded that the
halo is more likely to damp rather than excite bending waves.
However, this conclusion itself was called into question by \citet{binney1998} who carried out numerical experiments that incorporated a ring
model for the disc into a conventional $N$-body simulation of a dark
halo.  They found that when the disc is initialized to the
Sparke-Casertano tilt-mode, the warp rapidly winds up while its energy
remains constant or even increases.  \citet{binney1998} interpret
these results in terms of time-scales: the inner halo responds to the
disc on a time-scale comparable to the precession period and therefore
the potential that the disc precesses in will be markedly different
from the one assumed in the static-potential analysis of
\citet{sparke1984} and \citet{sparke1988}.  They go on to speculate
that there will be true modes of the disc-live halo system,
qualitatively similar to but quantitatively different from the
Sparke-Casertano modes, that nevertheless avoid the rapid damping
predicted in \citet{nelson1995}.

In this work, we show that a Fourier and
spectral analysis of bending waves yields a spectrum that is broadly
consistent with both the eigenmode and WKB analysis.  Moreover, there
is a low-frequency branch of modes with little differential
precession.  Waves along this branch are therefore long-lived and
thus candidates for the bending waves seen in the Galaxy.

Our work is based on $N$-body simulations of two
isolated Milky Way-like galaxy models, which are distinguished by the relative contributions
of the disc and halo to the rotation curve, and thus by their
susceptibility to the formation of a bar. The models and simulations are described in Section~\ref{sec:simulations}. In
Section~\ref{sec:bendingwaveanalysis}, we analyse the vertical bending
waves that emerge in our simulated galaxies.  In particular, we
decompose the waves according to their azimuthal symmetry and present
a spectral analysis that is reminiscent of the classic studies of
density waves \citep[see][]{sellwoodathanassoula1986}.  In
Section~\ref{sec:linearringmodel} we show that the qualitative
features seen in our spectral analysis are in good agreement with an
eigenmode analysis of the linear ring model as well as the dispersion
relation derived in the WKB approximation.
Section~\ref{sec:externalwaves} gives a brief description of how we
might observe vertical bending waves in external galaxies.  We discuss
our results in Section~\ref{sec:discussion} and conclude in
Section~\ref{sec:conclusions}.

\section{Simulations}
\label{sec:simulations}

\begin{figure} 
\includegraphics[width=\columnwidth]{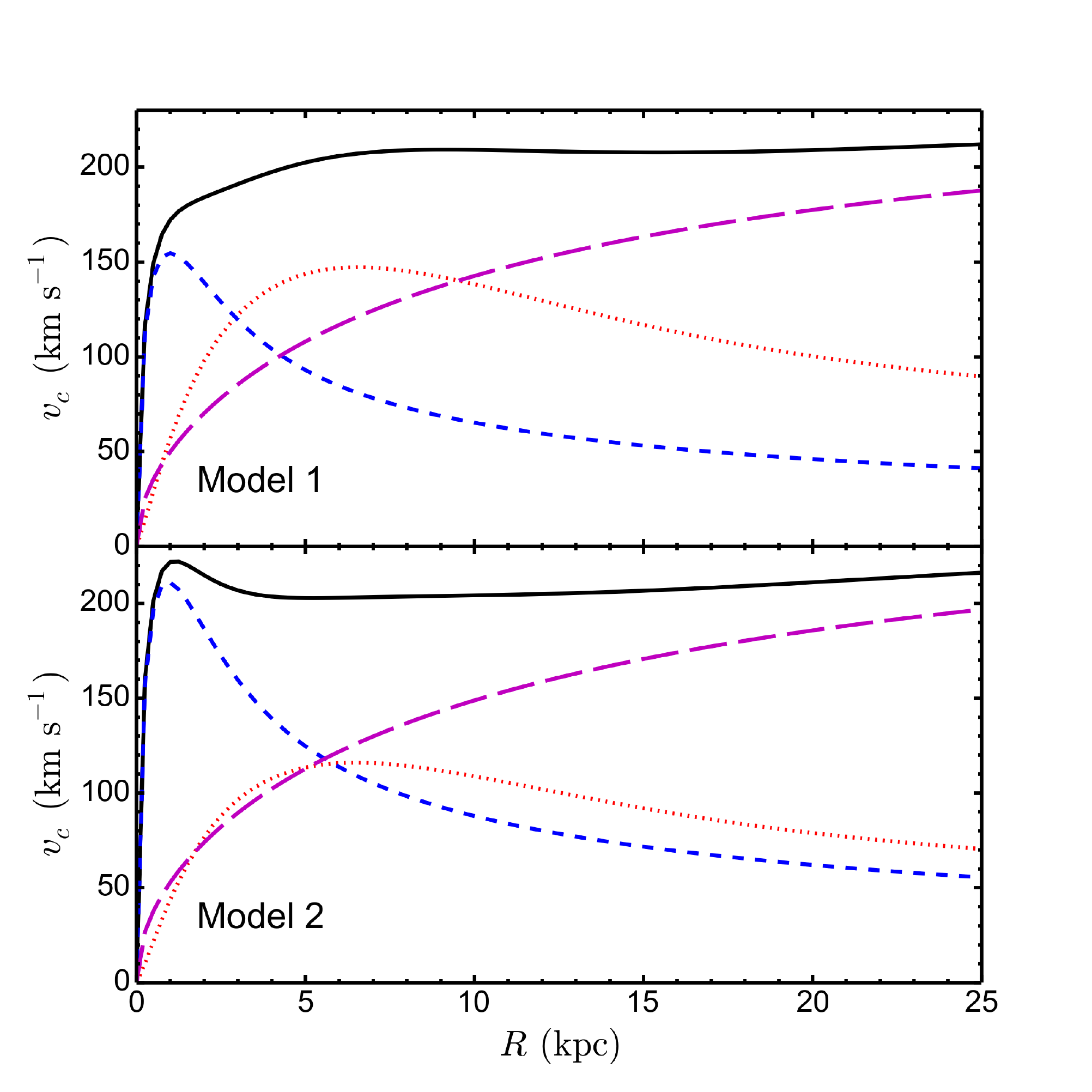}
\caption{Circular speed curves for the initial conditions of Model~1
  (top) and Model~2 (bottom).  The solid black curve shows the total
  circular speed of the model.  Also shown are the contributions to
  the circular speed curve from the bulge (blue, short-dashed), the
  disc (red, dotted), and the dark matter halo (magenta,
  long-dashed). Model~1 corresponds to the most stable Milky Way-like model
  of \citet{widrow2008}, and possesses a disc with enough self-gravity
  in the inner region to trigger bar formation. Model~2 was
  constructed with a disc mass 60\% of that in Model~1, while the
  bulge and halo parameters were adjusted to yield a similar total
  circular speed curve. Model~2 only forms flocculent spiral structure
  due to the dominant bulge and halo relative to the
  disc. \label{fig:circspeed}}
\end{figure}

\subsection{Two Milky Way-like Models}

We generate the initial conditions for our $N$-body simulations using
\textsc{galactics} \citep{kuijken1995, widrow2008}, which allows
one to construct axisymmetric disc-bulge-halo systems that are in
approximate dynamical equilibrium.  The phase space distribution
function (DF) for the disc in these models, $f_d$, is a function of
two exact integrals of motion, the energy and the component of the
angular momentum about the symmetry axis, and an approximate third
integral that corresponds to the vertical energy.  By construction,
the disc density obtained by integrating $f_d$ over all velocities is well-approximated by the function

\begin{equation} \label{eq:discdensity}
\rho_d\left (R,\,z\right ) = \frac{M_d}{4\pi R_d^2 h} e^{-R/R_d} {\rm
  sech}^2(z/h)\,C\left (\left (R-R_t\right )/\delta R_t\right ),
\end{equation}

\noindent where $R$ and $z$ are cylindrical coordinates and $C$ is a
truncation function that goes smoothly from unity to zero at $R\simeq R_t$
over a width of order $\delta R_t$.  The radial velocity dispersion is
an exponentially decreasing function of $R$:

\begin{equation}
\sigma_R(R) = \sigma_{R0} \exp{\left (-R/2R_d\right )},
\end{equation}

\noindent where we assume that the radial scale length for the square
of the velocity dispersion is the same as the scale length for the
surface density, in accord with observations \citep{bottema1993}.  The
azimuthal velocity dispersion profile is set by the epicycle
approximation (see \citealt{kuijken1995} and Section 3.2.3 of \citealt{binney2008}) while the
vertical velocity dispersion is determined from the local surface
density and the model assumption that the disc has a constant scale
height.

The DFs for the bulge and halo are functions solely of the energy and
are therefore isotropic in velocity space.  The bulge DF is
constructed to yield a density profile that is given, approximately,
by

\begin{equation} \label{eq:bulgedensity}
\rho_b\left (r\right ) = \frac{v_b^2}{4\pi G R_e^2 c(n)}
\left (\frac{r}{R_e}\right )^{-p}e^{-b\left (r/R_e\right )^{1/n}}~,
\end{equation}

\noindent which yields the S\'{e}rsic law for the projected surface
density profile with index $n$ so long as one sets $p=1-0.6097/n +
0.05563/n^2$.  The constant $b$ is adjusted so that $R_e$ encloses
half the total projected mass and $c(n) = \left (nb^{n\left
  (p-2\right)}\right )\Gamma\left (n\left (2-p\right )\right )$
\citep{prugniel1997,terzic2005}.

The halo DF is constructed to yield the NFW profile \citep{navarro1996}:

\begin{equation} \label{eq:halodensity}
\rho_h(r) = \frac{a_h v_h^2}{4\pi G}\frac{1}{r\left (r + a_h\right
  )^2}.
\end{equation}

\noindent While the velocity distributions of the halo and bulge are
isotropic, their space densities are slightly flattened due to the disc potential.

In this paper, we consider two \textsc{galactics} models.  The first (Model~1) is from
\citet{widrow2008}.  In that paper, Bayesian and Markov chain Monte Carlo methods
were used to constrain the \textsc{galactics} parameters so as to match kinematic and
photometric observations of the Milky Way.  That analysis yielded a suite of
models that varied in their susceptibility to bar formation.  The model that we
chose for this paper has a relatively weak bar instability.  For Model~2, we
reduced the disc mass by roughly a third and increased the masses of the bulge
and halo components so as to obtain a similar rotation curve.  The disc in Model~2 is therefore so light that it never forms a bar.

While our
models yield very similar circular speed curves they differ in the
relative contributions of the disc and the dynamically hot components
(the bulge and halo) to the radial force, as shown in
Fig.~\ref{fig:circspeed}. In particular, Model~1 has a disc mass of
$M_d \approx 4 \times 10^{10}\,M_\odot$, which impies that at a radius
of $2.2\,R_d$, $V_d^2/V_c^2 = 0.51$, where $V_c = V_c(R)$ is the
circular speed at radius $R$ and $V_d$ is the contribution to the
circular speed from the disc.  The central radial velocity dispersion
is set to $90\,{\rm km}\,s^{-1}$, which yields a Toomre $Q$ parameter
\citep{toomre1964} at $2.2\,R_d$ of $1.2$.  For Model~2, $V_d^2/V_c^2 = 0.33$ at
$2.2R_d$.  In addition, the central radial velocity dispersion in the
disc is decreased so that the Toomre $Q$ parameter at $2.2R_d$ is
the same as in Model~1.  The \textsc{galactics} parameters for the two
models are given in Table \ref{tab:initialmodelparameters}.

\begin{table}
\caption{Initial model parameters.}
\label{tab:initialmodelparameters}
\begin{threeparttable}
\begin{tabular}{lcc}
\hline
Parameter & Model~1 & Model~2\\
\hline
$M_d$ [$10^{10}\,M_\odot$]\tnote{a}  & 3.9  & 2.3 \\
$R_d$ [kpc]\tnote{b}   & 2.8  & 2.8 \\
$h$ [kpc]\tnote{c}      & 0.44  & 0.44 \\
$R_t$ [kpc]\tnote{d} & 25 & 25 \\
$\delta R_t$ [kpc]\tnote{e} & 3 & 3 \\
$\sigma_{R0}$ [${\rm km}\,{\rm s}^{-1}$]\tnote{f}      &  90  & 56.3  \\
$M_h$ [$10^{12}\,M_\odot$]\tnote{g}  &  1.2   &  1.3   \\
$v_h$ [${\rm km}\,{\rm s}^{-1}$]\tnote{h}   & 481 & 505 \\
$a_h$ [kpc]\tnote{i}   & 43 & 43 \\
$M_b$ [$10^{9}\,M_\odot$]\tnote{j}  &   9.9     &   17.9    \\
$v_b$ [${\rm km}\,{\rm s}^{-1}$]\tnote{k}  & 272 & 375 \\
$R_e$ [kpc]\tnote{l}   & 0.64 & 0.64 \\
$n$~\tnote{m}    & 1.32 & 1.32 \\
$Q(2.2R_{d})$~\tnote{n}   &  1.2    & 1.2    \\
\hline
\hline
\end{tabular}
\begin{tablenotes}
\item Disc mass\tnote{a}; radial disc scale length\tnote{b}; vertical disc scale height\tnote{c}; truncation radius of the disc\tnote{d}; truncation width of the disc\tnote{e}; central radial velocity dispersion in the disc\tnote{f}; halo mass\tnote{g}; characteristic velocity scale of the halo\tnote{h}; NFW halo scale length\tnote{i}; bulge mass\tnote{j}; characteristic velocity scale of the bulge\tnote{k}; bulge effective radius\tnote{l}; S\'{e}rsic index\tnote{m}; Toomre $Q$ parameter at 2.2$R_{d}$~\tnote{n}.
\end{tablenotes}
\end{threeparttable}
\end{table}

\subsection{\boldmath{$N$}-body Simulations}
\label{sec:nbodysimulations}

\begin{figure*}
\includegraphics[width=\linewidth]{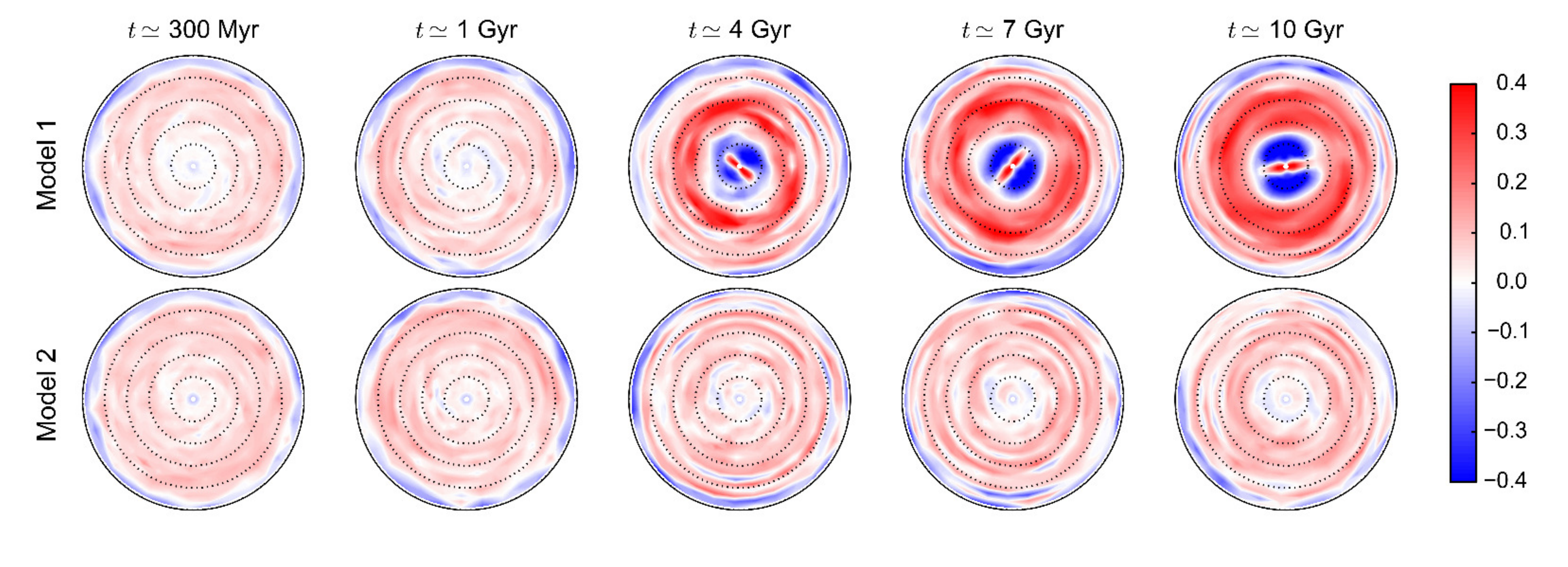}
\caption{Face-on disc surface density enhancement for Model~1 (top row) and Model~2 (bottom row) at five epochs throughout the simulation, as indicated at the top of each column. The colour map indicates the logarithm (base 10) of the ratio of the surface density at each epoch relative to the initial equilibrium surface density. Dotted concentric circles indicate increments of 5 kpc in radius. The rotation of the discs is counter-clockwise. The bar that forms in the higher mass disc of Model~1 is clearly visible, as well as a ring structure between 10 and 15~kpc. Model~2 only forms flocculent tightly wound spiral structure and displays only modest surface density enhancements relative to Model~1. \label{fig:surfacedensity}}
\end{figure*}

\begin{figure*}
\includegraphics[width=\linewidth]{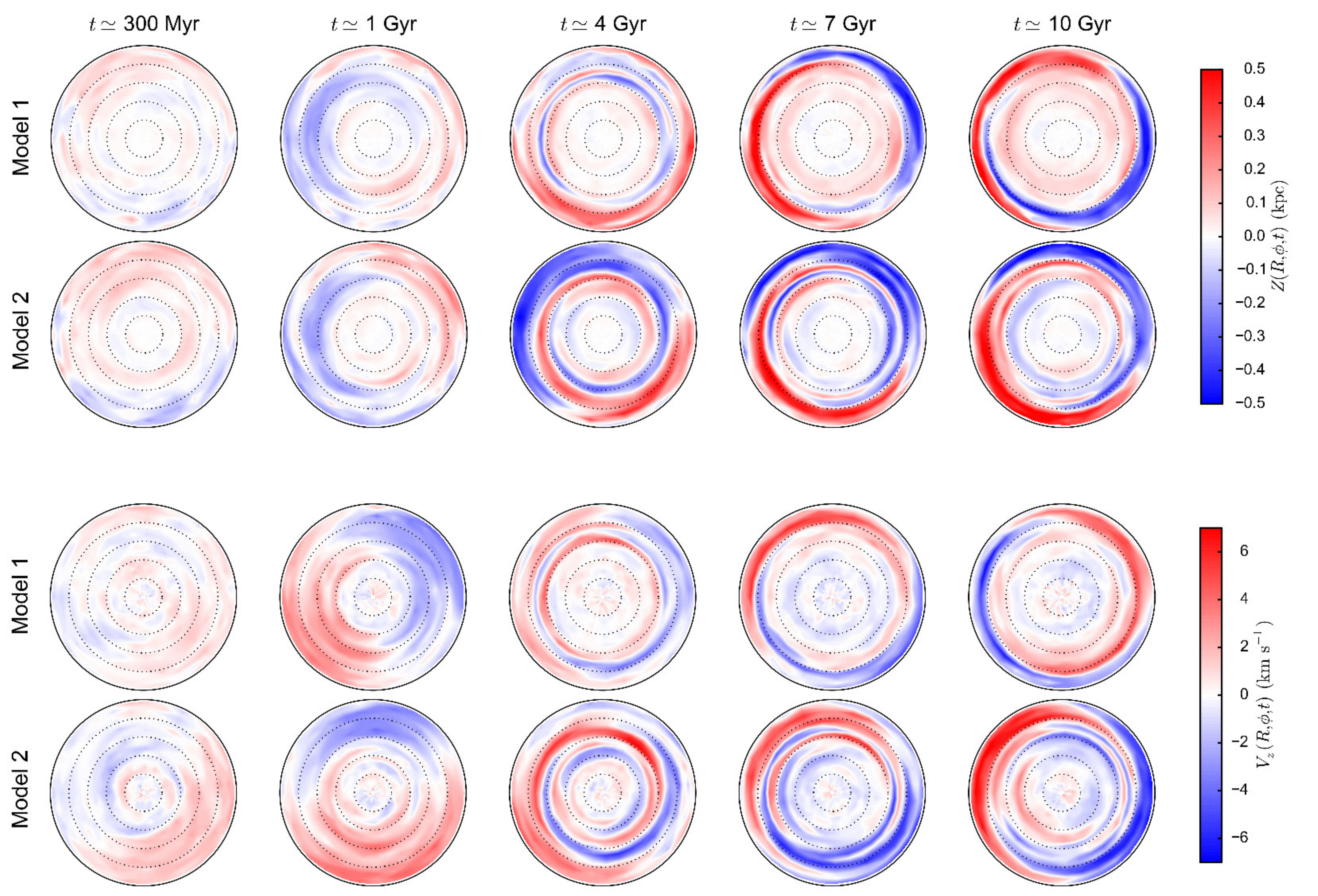}
\caption{Face-on maps of the mean vertical displacement, $Z\left (R,\,\phi,\,t\right )$ (top set of rows), and mean vertical velocity, $V_z\left (R,\,\phi,\,t\right )$ (bottom set of rows), for Model~1 and Model~2, as indicated on the left side. Columns correspond to the same five epochs in Fig.~\ref{fig:surfacedensity}, as indicated at the top of each column. Dotted concentric circles indicate increments of 5 kpc in radius. The rotation of the discs is counter-clockwise. At later times, the leading nature of the bending waves, especially in the outer disc, is very clear. In general, bending is slightly stronger and more pervasive in the less massive and barless disc of Model~2 compared to Model~1, especially at later epochs. Furthermore, the average height and vertical velocity seem to be coupled, qualitatively similar to the oscillatory anti-correlation previously reported \citep{gomez2013,gomez2016,gomez2017}.
 \label{fig:zmap_vmap_bar_nobar}}
\end{figure*}

We use \textsc{galactics} to generate $N$-body realizations for the two
models just described. These realizations have $2.5M$ disc particles,
$250k$ bulge particles, and $5M$ halo particles. The masses of particles within each component are identical. The initial
conditions were evolved for $\sim\!10\,{\rm Gyr}$ using
\textsc{gadget-3} \citep{springel2005} with a softening length of
$40\,{\rm pc}$ for all particle types. The maximum time step was set
to $0.2\,{\rm Myr}$, which is $\sim$~0.2\% of the Galactic dynamical
time at a radius of 20 kpc, where the warp occurs, and $\sim$~1\% of a Galactic dynamical time at 4 kpc.  Total energy was conserved to within
0.06\%. We start with ``pristine" equilibrium initial conditions, which allow
us to focus on key dynamical processes such as the secular evolution
of bending waves.  Our simulations ignore gas dynamics and
do not capture the physics of galaxy formation.

Over time the discs in our simulations rotate and drift slightly about
the global coordinate origin. To account for this when analyzing
face-on maps of our discs we center the
disc and rotate it into the $x$-$y$ plane by way of the following iterative scheme.  We compute the center of mass of disc particles
within a cylinder of radius 20 kpc and height 2 kpc centered in the
$x$-$y$ plane and translate all disc particles to that frame. This procedure is
repeated until a translation of less than 1 pc is achieved in all
coordinate directions. Next, we account for any tilt of the disc
relative to the $x$-$y$ plane by using a two-dimensional
Newton-Raphson scheme to find the Euler angles about the $x$- and $y$-
axes that minimize the root mean square vertical displacement of disc
particles within a cylinder of radius 20 kpc and height 2 kpc centered
in the $x$-$y$ plane. This method does a good job of diagonalizing the
moment of inertia tensor of the disc.

In Fig.~\ref{fig:surfacedensity}, we show the disc surface density relative to the
equilibrium surface density for the two models at various times.  The
maps extend to $R=25 \, {\rm kpc}$, or about $9 R_d$, which is the radius at which
we begin to truncate the disc.  This radius also corresponds to the
outermost radius at which the warp in the Milky Way's stellar disc has
been observed \citep{momany2006}.  The decrease of $\sim 20 - 55\%$ in the surface density
for $R  > 22 \, {\rm kpc}$, which develops at early times, is likely due to a relaxation
in the initial conditions.

The disc in Model~1 forms a bar at $\sim\!2.5\,{\rm Gyr}$, which eventually extends
to $R \sim 5 \, {\rm kpc}$.  In addition, a prominent ring (again, an enhancement
relative to the initial exponential profile) develops at a radius
of $\sim 10 - 15\,{\rm kpc}$.  This feature is likely a result of mass redistribution
due to the bar.  On the other hand, the disc in Model~2 forms
flocculent spiral structure but no bar.

\section{Analysis of Bending Waves}
\label{sec:bendingwaveanalysis}

\subsection{Fourier Analysis}
\label{sec:fourieranalysis}

The surface density shown in Fig.~\ref{fig:surfacedensity} represents
the zeroth moment, with respect to $z$, of the coarse-grained stellar
DF, $\tilde f$: 

\begin{equation}
\Sigma\left (R,\,\phi,\,t\right ) = \int \tilde{f}\left ({\bf r},\,{\bf v},\,t\right
) d^3\!v \,dz~.
\end{equation}

\noindent (The true DF for an $N$-body system is a sum of
six-dimensional $\delta$-functions at the positions of the particles.
In practice, coarse-graining amounts to calculating an average surface
density over a small patch of the disc.)  Likewise, the mean vertical
displacement of the disc is given by the first moment with respect to $z$:

\begin{equation}\label{eq:zdisplacement}
  Z\left (R,\,\phi,\,t\right ) = \Sigma\left (R,\,\phi,\,t\right )^{-1}
  \int \tilde{f}\left ({\bf r},\,{\bf v},\,t\right
  ) \, z \, d^3\!v \,dz~.
\end{equation}

In Fig.~\ref{fig:zmap_vmap_bar_nobar}, we show mean vertical displacement and vertical bulk
velocity maps for the two models at the same five epochs shown in
Fig.~\ref{fig:surfacedensity}.  During the first Gyr, displacements of order 100 pc and bulk
vertical motions of a few ${\rm km\,s}^{-1}$ are found across the discs in both
models.  We interpret these displacements and bulk motions as bending
waves.  By 4 Gyr, the presence of the bar is clearly evident in Model~1.  Interestingly enough, by this epoch, the vertical displacements
are stronger in Model~2 and extend further in toward the centre of the
disc, as compared with Model~1.  The implication is that the bar
and/or more massive nature of the disc in Model~1 has a damping effect
on bending waves.

The waves have a strong $m=1$ component, where $m$ is the azimuthal mode
number.  Moreover, these waves appear to be leading in the sense that
the radius of the peak vertical displacement ridge increases with
increasing $\phi$, that is, in the direction of rotation.  The $m=1$ waves
are tightly wound at intermediate radii but transition smoothly to a
warp at the edge of the disc.  In general, the amplitude of the waves
in the outer disc grows over time while the amplitude decreases in the
inner disc.  We return to this point below.

The bulk vertical motions follow the same general pattern of the displacements
except that they are out of phase, as one would expect for wave-like motion \citep[see][]{gomez2013,gomez2016,gomez2017}. From the amplitudes of the velocity and height perturbations, we infer a characteristic pattern speed for the bending waves of $\sim 10-20\,{\rm km\,s}^{-1}\,{\rm kpc}^{-1}$.

In the study of density waves, it has proved instructive to write the
surface density as a Fourier series in $\phi$ \citep[see, for example, the seminal work of][]{sellwoodathanassoula1986}.  Formally, one has

\begin{equation} \label{eq:surfdens}
\Sigma\left (R,\,\phi,\,t\right ) = {\rm Re}\left \{ \sum_m \Sigma_m\left (R,\,t\right
) e^{-im\phi}\right \}~,
\end{equation}

\noindent where

\begin{equation}
\Sigma_m\left (R,\,t\right ) = \frac{1}{2\pi}\int_0^{2\pi}
\Sigma\left (R,\,\phi,\,t\right
) e^{im\phi} \, d \phi~.
\end{equation}
 
\noindent In practice, the $\Sigma_m$ are calculated over a ring of
finite radius (essentially, coarse graining in $R$).  For a ring
at radius $R_\alpha$ and width $\Delta R_\alpha$ we have

\begin{equation}
\Sigma_m\left (R,\,t\right ) = A_\alpha^{-1} \sum_{j\in\alpha} m_j \,
e^{im\phi_j}~,
\end{equation}

\noindent where $A_\alpha = 2\pi R_\alpha \Delta R_\alpha$ is the area
of the ring and the sum is over all particles in the ring.  Similar to
the surface density, we can construct a Fourier series for the
vertical displacement:

\begin{equation}\label{eq:displacement}
Z\left  (R,\,\phi,\,t\right ) = {\rm Re}\left \{ \sum_{m=0}^\infty
Z_m\left (R,\,t\right ) e^{-im\phi}\right \}~,
\end{equation}

\noindent where we define

\begin{equation} \label{eq:bendmodefouriercoef}
Z_m\left (R,\,t\right ) \equiv A_\alpha^{-1} \sum_{j\in\alpha} 
\frac{z_j \, m_j }{\Sigma\left (R_j,\,\phi_j,\,t\right )} \, e^{im\phi_j}~.
\end{equation}

\noindent The surface density at particle $j$, $\Sigma\left
  (R_j,\,\phi_j,\,t\right )$, is estimated by using the Fourier series
for $\Sigma$ (equation~\ref{eq:surfdens}) truncated at the 5th order terms.

\begin{figure}
\includegraphics[width=\columnwidth]{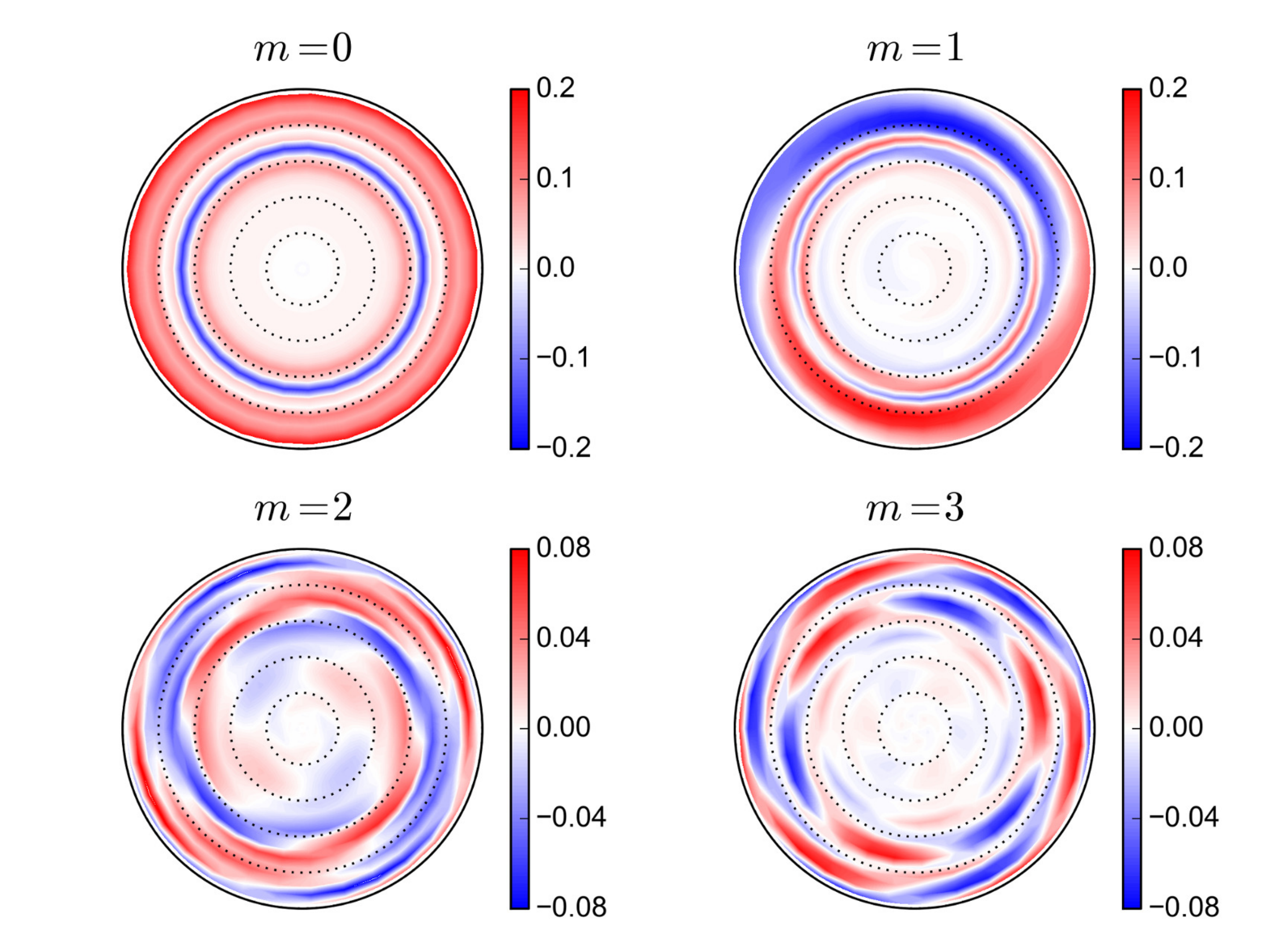}
\caption{Fourier mode decomposition of $Z(R, \phi, t \sim 4\,{\rm Gyr})$ for $m=0,\,1,\,2,$ and $3$, as indicated, for Model~1. Dotted concentric circles indicate increments of 5 kpc in radius. The rotation of the disc is counter-clockwise. The colour scale is in units of kpc, and differs between the upper and lower panels in order to highlight the extremal values for each $m$. The factor of $1/2$ for the zeroth order term in the Fourier Series is reflected in the $m = 0$ panel. From the figure it is clear that the (leading) $m = 1$ term is the dominant one in the decomposition.
\label{fig:Zm_maps}}
\end{figure}

In Fig.~\ref{fig:Zm_maps} we show the $m=0,\,1,\,2,$ and $3$ Fourier
modes of $Z$ for the $t \sim 4\,{\rm Gyr}$ snapshot of the Model~1
simulation.  We see that while the $m=1$ mode is clearly leading, the
$m=2$ term is trailing, and the $m=3$ term is a combination of both
leading and trailing. Furthermore, the $m=1$ mode is the dominant mode
beyond a radius of $\sim 15 \, {\rm kpc}$.

\begin{figure}
\includegraphics[width=\columnwidth]{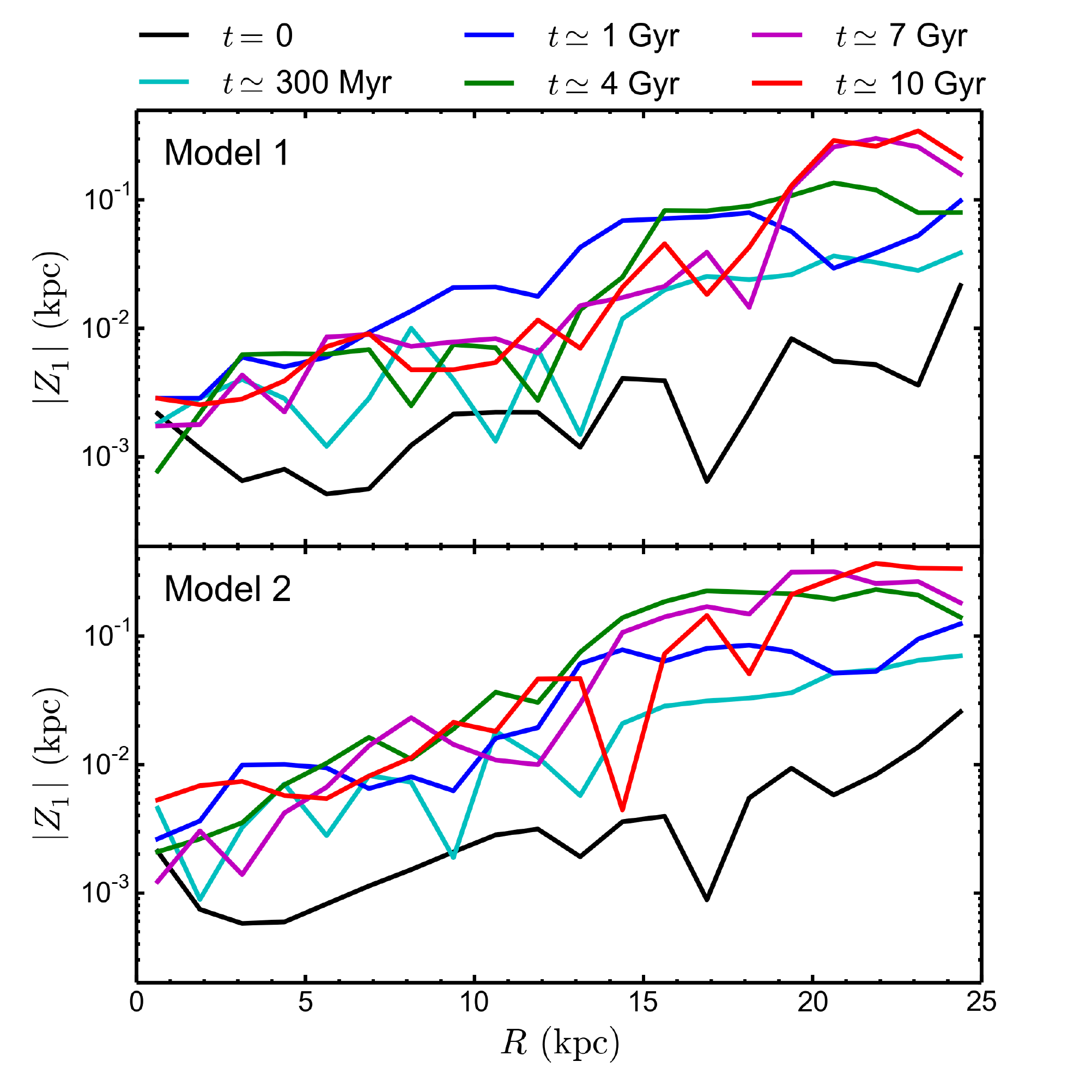}
\caption{Time evolution of $|Z_1(R)|$. The upper and lower panels show Model~1 and Model~2, respectively. Both models show a general trend of increasing strength propagating outward from intermediate radii to the outer disc over time.
\label{fig:PS_evolution}}
\end{figure}

We explore how this dominant mode in the outer disc forms and evolves
in Fig.~\ref{fig:PS_evolution} by showing
$|Z_1(R)|$ for both models at various snapshots. Both models show a trend of increasing $|Z_{1}(R)|$ from intermediate radii to the outer disc with time. This is most evident in Model~1 where there is a factor of $\sim 2-3$ decrease in $|Z_1|$  at $R \sim 10 - 15 \, {\rm kpc}$ and an approximately equal increase in the outer disc after $t \simeq 1$~Gyr. We attribute the large suppression of $|Z_1|$ at intermediate radii to the bar, which forms at $\sim 2.5$~Gyr. This trend, which is also present in Model~2 though not so pronounced, suggests that warps may arise from the outward migration of $m=1$ bending waves.

\subsection{Spectral Analysis}
\label{sec:spectralanalysis}

To connect our 3D simulations with the linear analysis of
Section~\ref{sec:linearringmodel}, we perform a spectral analysis of
bending waves in our simulated discs.  Our method uses the techniques
from \citet{sellwoodathanassoula1986} (see also \citealt{roskar2012} for a succinct description).  We calculate the
Fourier coefficients $Z_m (R,t)$ (see
Section~\ref{sec:fourieranalysis},
equation~\ref{eq:bendmodefouriercoef}) for the $t_j=j \, \Delta_t +t_0$
snapshots where $\Delta_t$ is the time between snapshots, $j=0\dots N-1$, and $N$ is even.  We then perform a
discrete Fourier transform in time to obtain the two-sided spectral coefficients

\begin{equation} \label{eq:freqfouriercoefs}
\mathcal{Z}_{m} (R,\omega_k) = \sum_{j = 0}^{N-1} Z_{m} (R, t_j) \, w(j) \, e^{2 \pi i j k/N}~,
\end{equation}

\noindent \citep[][Section 13.4]{press2007}. Here, $w(j)$ is a Gaussian window function with a standard deviation of $N/2^{5/2}$, which is introduced to mitigate spectral leakage from high frequencies. We note that using similarly shaped window functions, such as the Hamming window, does not significantly alter the resulting spectra or our conclusions. The discrete frequencies\footnote{Since we consider
negative frequencies, we note a difference between our index $k$,
and that used in the conventional spectral analysis. For example,
the mapping between our $k = -N/2\dots N/2$ and the index used in
\citet{roskar2012}, $k' = 0 \dots N-1$, is $k' = k$ for $k \ge 0$
and $k' = k + N$ for $k < 0$. Therefore, the range $k' = 0 \dots N/2
-1$ corresponds to increasing positive frequencies, and $k' = N/2 +
1 \dots N-1$ corresponds to increasing negative frequencies. The Nyquist frequency for $k' = N/2$ corresponds to $k = \pm
N/2$.} are given by

\begin{equation}
\omega_k = \frac{2 \pi}{m} \frac{k}{N \Delta_t}~,
\end{equation}

\noindent with $k = -N/2\dots N/2$, and the Nyquist frequency corresponding to $\omega_{k= \pm N/2}$. The two-sided power spectrum is then computed as

\begin{equation}
P_m (R,\omega_k) = \frac{1}{W} \, |\mathcal{Z}_m (R,\omega_k)|^2 ~,
\end{equation}

\noindent where $W = N \, \sum_{j = 0}^{N-1} w(j)$ is the window
function normalization.

\begin{figure*}
\includegraphics[width=\linewidth]{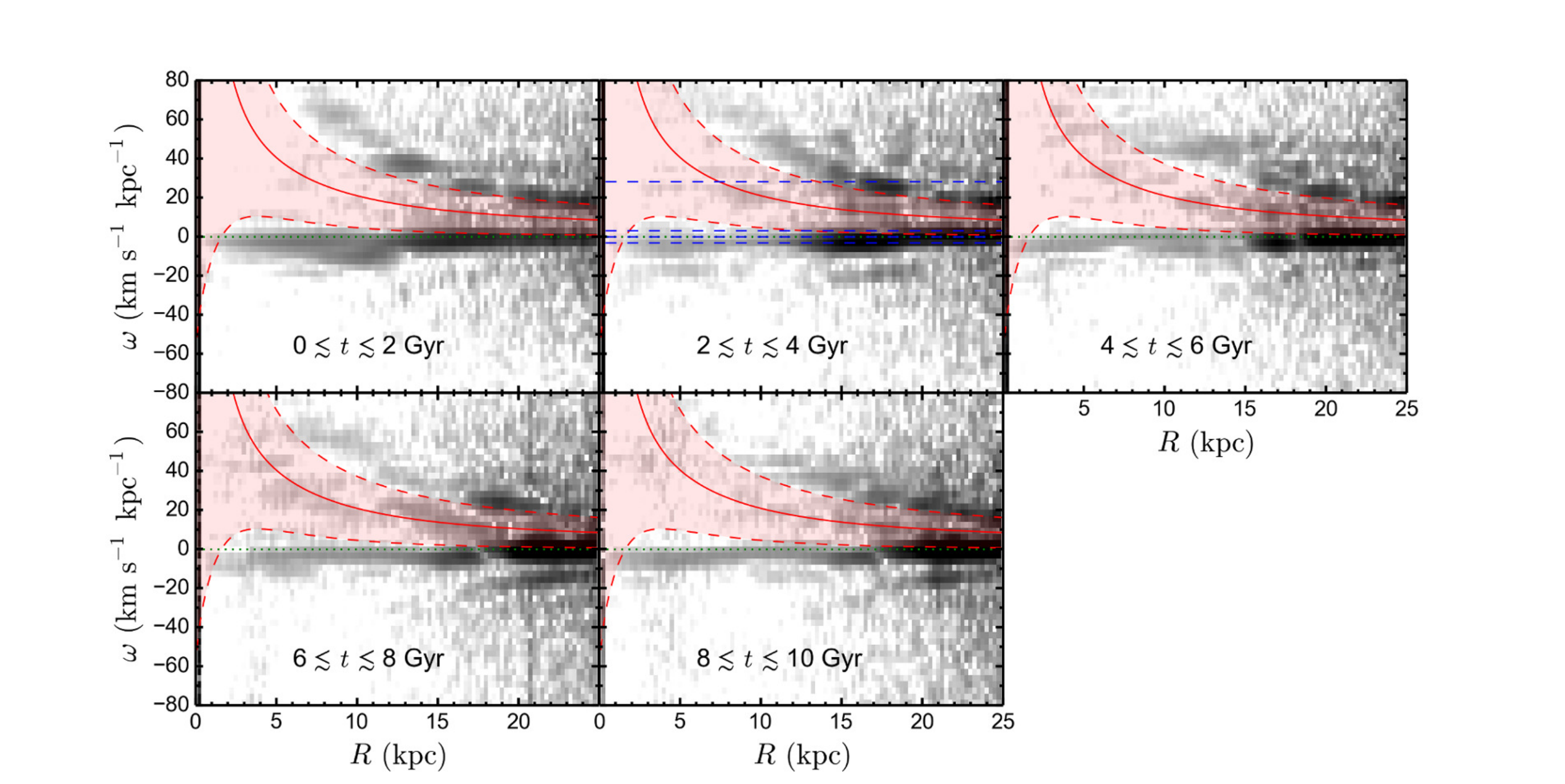}
\caption{Two-sided frequency power spectra of $m=1$ bending waves for various $\sim 2 \, {\rm Gyr}$ time intervals, as indicated, for Model~1. The scale of power is logarithmic and is constant across all time intervals shown. Overlaid on all panels in red are total circular frequency (solid) and vertical resonance (dashed) curves, $\omega = \Omega \pm \nu_x$, where $\nu_x = \sqrt{\nu_{b}^{2} + \nu_{h}^{2}}$ is the vertical forcing frequency from the bulge and halo, for the equilibrium initial conditions. The shaded red indicates the ``forbidden" region between the resonances according to the WKB approximation (see Section 6.6.1 of \citealt{binney2008} and Section~\ref{sec:linearringmodel}). The horizontal dotted green line references zero frequency. The horizontal dashed blue lines in the middle top panel indicate the frequencies considered in Fig.~\ref{fig:tip-LON_2_to_4Gyr}.
  \label{fig:powertimeseries}}
\end{figure*}

In Fig.~\ref{fig:powertimeseries}, we explore the time evolution of the $m=1$ bending wave power
spectra for Model~1.  To do so, we divide the simulation into five
$\sim 2 \, {\rm Gyr}$ intervals and calculate $P_1$ using the method outlined above.
Each interval has $N=200$ snapshots with a time resolution of $\Delta_t \approx 9.8 \, {\rm Myr}$, which gives a frequency
resolution of $\sim 3.14 \, {\rm km \, s^{-1} \, kpc^{-1}}$.
Prominent bands of power develop at early times and persist over the entire simulation.  Much of the power lies along two main branches, which, as
we describe below, follow the two lowest-order vertical resonances
from linear theory.  Power along the lower branch, which correponds to
slowly counter-rotating waves, is seen across the disc at all times.
On the other hand, power along the upper branch (more rapidly rotating
waves) diminshes with time.  In general, over time power migrates to
larger radii.

We note the presence of spurious bands of power at fixed radii that extend
almost the entire frequency range in
Fig.~\ref{fig:powertimeseries}. More specifically, there is a band
present in all time intervals at $R < 0.25 \, {\rm kpc}$, as well as
weaker bands at larger radii that come and go across the various time
intervals. These bands arise when the local surface density term in the vertical displacement Fourier coefficients ($\Sigma\left (R_j,\,\phi_j\right )$ in equation~\ref{eq:bendmodefouriercoef}) for one or more particles is
computed to be nearly zero.  This situation is an artefact of our
scheme to estimate the local surface density as a truncated Fourier
series.

Overlaid in all panels of Fig.~\ref{fig:powertimeseries} are the $m =1$ vertical resonances (analogous to the Lindblad resonances for
density waves), $\omega = \Omega \pm \nu_x$, where $\Omega$ is the
circular frequency of the disc and $\nu_x = \sqrt{\nu_{b}^{2} +
  \nu_{h}^{2}}$ is the vertical force oscillation frequency due to the
bulge and halo potential, of the initial conditions. Power in the $m =1$ bending waves collects along the vertical resonances and tends to lie outside of the forbidden region between them, as predicted by the WKB approximation \citep[][Section 6.6.1, but also see Section~\ref{sec:linearringmodel}]{binney2008}. Furthermore, agreement between our power spectra and the vertical resonances of the initial conditions endures for many billions of years.

We can then ask the question as to why higher frequency waves subside
over time, or conversely, why low-frequency waves persist?  To this end, we consider a form of the tip-Line-of-Nodes (LON) plot
\citep[see][]{briggs1990} for waves of various pattern speeds.  A
tip-LON plot is constructed by calculating, for a series of concentric
rings, the tilt, or tip away from the midplane, of the disc and the
position of the LON.   To do this, we construct an effective
vertical displacement as a function of $R$ and $\phi$ for a narrow range in frequency, $\Delta\omega_k$, using only the $m=1$
spectral coefficient of the $t - \omega_k$ Fourier transform:

\begin{equation}
Z_1\left (R,\,\phi,\,\omega_k\right ) = 
\int_{\omega_k - \Delta\omega_k/2}^{\omega_k+\Delta\omega_k/2} {\rm Re} \bigg\{ \mathcal{Z}_1\left
  (R,\,\omega_k\right )e^{i\phi} \bigg\} \, d\omega_k~.
\end{equation}

\noindent The function $Z_1\left (R,\,\phi,\,\omega_k\right )$ is
essentially our series of rings from which the tilt and position angle for the LON as a function of $R$ can be easily determined.

\begin{figure}
\includegraphics[width=\columnwidth]{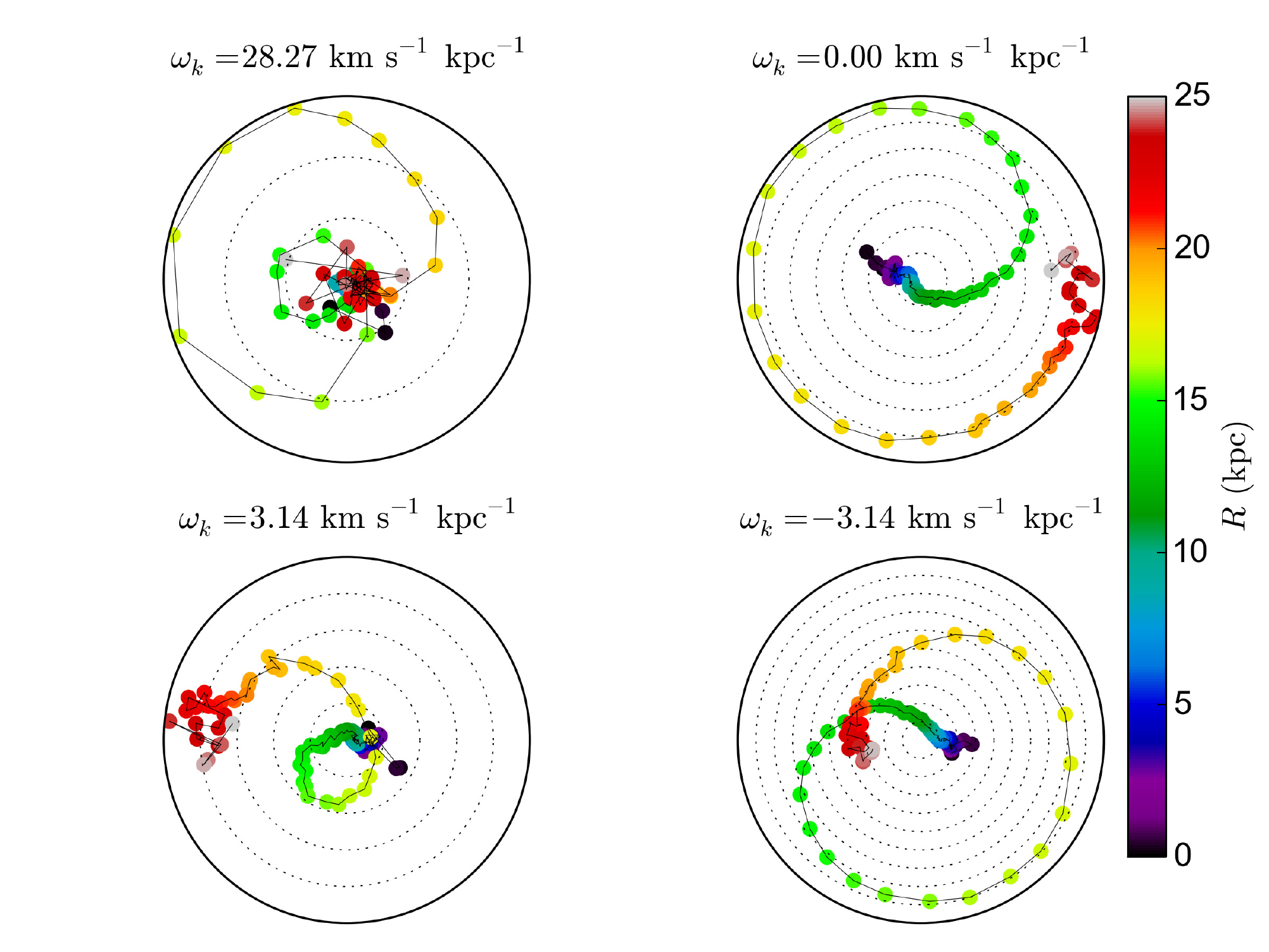}
\caption{A form of tip-LON plots \citep[so-called ``Briggs" plots,
  see][]{briggs1990} for four wave frequencies within the time
  interval $2 \la t \la 4 \, {\rm Gyr}$ (indicated in
  Fig.~\ref{fig:powertimeseries}, top middle panel). For each
  frequency the disc is divided into concentric rings in radius. Each
  point in the figure corresponds to a radial ring, with the radius
  indicated by colour. The radial coordinate is the tilt, in degrees,
  of that ring, and the azimuthal coordinate is the position of
  maximum vertical displacement. Concentric dotted circles in the radial direction indicate tilt
  increments of $5\degr$. High-frequency waves are morphologically trailing since the azimuth decreases with increasing radius. In contrast, waves near zero frequency are leading.
  \label{fig:tip-LON_2_to_4Gyr}}
\end{figure}

Tip-LON plots are shown in Fig~\ref{fig:tip-LON_2_to_4Gyr}.  For illustrative purposes we only consider the time interval $2 \la t \la 4 \, {\rm Gyr}$ for Model~1 (i.e. Fig.~\ref{fig:powertimeseries}, top middle panel) and show the results for $\omega_k = 28.3, \, 3.1, \, 0,$ and $-3.1 \, {\rm km \, s^{-1} \, kpc^{-1}}$.  These four frequencies, which are indicated by
horizontal dashed lines in Fig.~\ref{fig:powertimeseries}, correspond to a typical frequency in the
upper branch and three frequencies in the lower branch.  Each dot of a tip-LON plot
corresponds to a ring within the disc whose radius, tilt, and position
angle are encoded in the colour, radial coordinate, and angular
position, respectively, of the dot.  Thus, dots that trace a clockwise
pattern with increasing ring radius indicate a trailing wave.

Large tilt angles for $\omega_k = 28.3 \, {\rm km \, s^{-1} \, kpc^{-1}}$ are localized to $15 \lesssim R \lesssim 20 \, {\rm kpc}$
where there is a prominent peak in the power spectrum (see Fig.~\ref{fig:powertimeseries}).
These are clearly trailing waves.  On the other hand, the tilt is
stronger across a greater range in $R$ at $\omega_k = -3.1, \, 0,$ and $3.1 \, {\rm km \, s^{-1} \, kpc^{-1}}$.  These
are leading waves, which experience less shear in the outer disc and therefore persist over a longer period of time.

\section{Linear Ring Model}
\label{sec:linearringmodel}

In this section, we study the linear bending waves of a razor thin,
self-gravitating disc embedded in the static gravitational potential
of a bulge and halo. The starting point for this investigation is the
second-order integro-differential equation for local vertical
displacements of the disc, $Z(R, \, \phi, \, t)$, which was introduced
by \citet{hunter1969}.  In order to relate this $Z$ to the mean
vertical displacement mapped in our simulations, we first show that
the equation from \citet{hunter1969} can be derived from the
collisionless Boltzmann equation.

\subsection{Equations of Motion}

We consider a system of collisionless ``stars'' whose equilibrium
distribution is that of a self-gravitating disc embedded in the
external potential of an extended dark matter halo and a centrally
concentrated bulge.  We assume that the stellar orbits, when projected onto the disc plane, are circular.  The DF for the stars can then be written

\begin{equation} 
f({\bf r},{\bf v},t) = 
\delta(v_R) \, \delta(v_\phi -R \, \, \Omega(R)) \,F\rpzvt~,
\end{equation}

\noindent where $\Omega(R) = v_c(R)/R$, $\delta(\cdot)$ is the Dirac delta function, and $F\rpzvt$ is the reduced DF.  Upon integration over $v_R$
and $v_\phi$, the collisionless Boltzmann equation becomes

\begin{equation}\label{eq:reducedCBE}
\frac{\partial F}{\partial t} + \Omega(R)
\frac{\partial F}{\partial\phi}  +
v_z\frac{\partial F}{\partial z}
= 
\frac{\partial \psi}{\partial z}\frac{\partial F}{\partial v_z} ~,
\end{equation}

\noindent where $\psi$ is the gravitational potential.

The vertical displacement of the disc is found by taking the 
$z$-moment of the DF:

\begin{equation}
Z(R, \, \phi, \, t) = \int \,z\,F\rpzvt\,dz\,dv_z~.
\end{equation}

\noindent This quantity is essentially the analog of $Z$ in 
equation~(\ref{eq:zdisplacement}) for the thin, cold disc. Likewise, the vertical motion of the disc is

\begin{equation}
V(R, \, \phi, \, t) = \int v_z\,F\rpzvt\,dz\,dv_z~.
\end{equation}

\noindent Multiply equation~(\ref{eq:reducedCBE}) by either $z$ or $v_z$ and
integrate over $z$ and $v_z$ and we obtain

\begin{equation}
\left (\frac{\partial}{\partial t} + \Omega(R)
\frac{\partial }{\partial \phi}\right ) Z
= V
\end{equation}

\noindent and

\begin{equation}
\left (\frac{\partial}{\partial t} + \Omega(R)
\frac{\partial }{\partial \phi}\right ) V
= -\frac{\partial \psi}{\partial z} = F_b + F_h + F_d ~,
\end{equation}

\noindent where in the last line, we've written out the separate
contributions to the vertical force from the bulge, halo, and disc.
Together, these equations lead to the standard second order equation
for $Z$, which was first introduced in \citet{hunter1969}.

For definiteness, we assume that the potential from the bulge and halo
are the same as in the \textsc{galactics} model described in Section
2.  Near $z=0$, the contributions to the vertical restoring force due
to these components are given, to a good approximation, by

\begin{equation}
F_b \left ({\bf R},\,t\right ) = -\nu_b^2 \, Z \left ({\bf R},\,t\right )~~~~~~~~~~
F_h \left ({\bf R},\,t\right ) = -\nu_h^2 \,  Z \left ({\bf R},\,t\right )~,
\end{equation}

\noindent where ${\bf R} = (R,\phi)$ in cylindrical coordinates. The vertical force due to the disc, to first order
in $Z/R$, is

\begin{equation}\label{eq:discforce}
{\bf F}_d = -G \int d{\bf R'} \Sigma(R')
\frac{Z({\bf R},\,t) - Z({\bf R}',t)}
{\left (R^2 + R'^2 - 2RR'\cos{\varphi} + z_0^2\right )^{3/2}}~,
\end{equation}

\noindent where $\varphi \equiv \phi - \phi'$ and the ``softening"
parameter $z_0$ is introduced to make the integral regular at ${\bf
  R}={\bf R}'$ \citep{sparke1988}.  (See \citealt{hunter1969},
\citealt{sparke1988}, and \citealt{nelson1995} for alternative ways of handling the ${\bf R}={\bf R}'$ singularity in the integrand.)  Note that the term on the right-hand side that is proportional to $Z \left ({\bf R},\,t\right )$ may be written $F_1 = -\nu_d^2 \, Z$, where

\begin{equation}
\nu_d^2 = G\int_0^\infty dR' R' \Sigma(R') 
H\left  (R,\,R'\right )
\end{equation}

\noindent and

\begin{equation}
H\left (R,\,R'\right ) = \int_0^{2\pi}\frac{d\varphi}
{\left (R^2 + R'^2 - 2RR'\cos{\varphi} + z_0^2\right )^{3/2}} ~.
\end{equation}

\noindent This term describes the vertical restoring force due to the
equilibrium disc that acts on an element of the disc
displaced from the midplane.  We can then write

\begin{equation}
\nu_t^2 = \nu_d^2 + \nu_b^2 + \nu_h^2 ~,
\end{equation}

\noindent where $\nu_t$ is the total vertical epicyclic frequency.  The term
in equation~(\ref{eq:discforce}) that is proportional to $Z \left ({\bf R'},\,t\right )$
describes the perturbing force on an unperturbed element in the disc
due to those parts of the disc that have been displaced from the
midplane.

For small displacements, we can write $Z$ and $V$ as superpositions of
linear modes with $m$-fold azimuthal symmetry so that each mode has
the form 

\begin{equation}
Z(R, \, \phi, \, t) = {\rm Re}
\left \{{\cal Z}(R,\,\omega) e^{i\left ( m\phi-\omega t\right  )}\right \}
\end{equation}

\noindent and

\begin{equation}
V(R, \, \phi, \, t) = {\rm Re}
\left \{{\cal V}(R,\,\omega) e^{i\left (m\phi-\omega t\right  )}\right \}~.
\end{equation}

\noindent We then have

\begin{equation}\label{eq:dispersion1}
-i\left (\omega - m\Omega\right ) {\cal Z}(R,\,\omega) = {\cal V}(R,\,\omega)
\end{equation}

\noindent and

\begin{equation}
\begin{aligned}
\label{eq:dispersion2}
-i\left (\omega - m\Omega\right ) & \, {\cal V}(R,\,\omega) = -\nu_t^2 {\cal Z}(R,\,\omega) \\
& + G\int_0^\infty
dR' R' \Sigma(R') {\cal Z}(R',\,\omega) I_m(R,\,R')~,
\end{aligned}
\end{equation}

\noindent where

\begin{equation}
I_m\left (R,\,R'\right ) = \int_0^{2\pi}\frac{d\varphi\cos{m\varphi}}
{\left (R^2 + R'^2 - 2RR'\cos{\varphi} + z_0^2\right )^{3/2}} ~.
\end{equation}

\subsection{Linear Ring Model and the WKB Approximation}
\label{sec:linearringandWKB}

\begin{figure}
\includegraphics[width=\columnwidth]{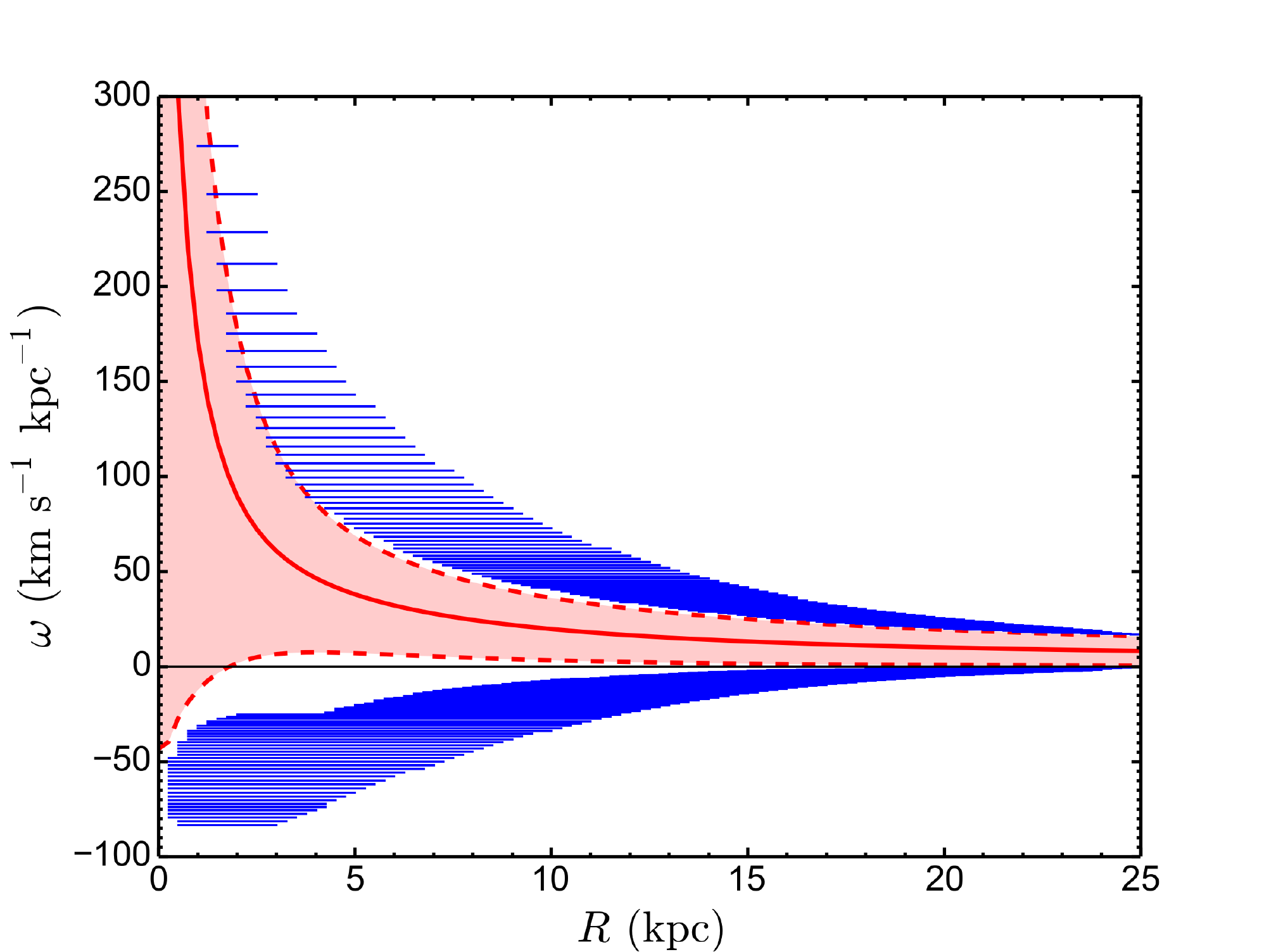}
\caption{Frequency as a function of radius for $m=1$ bending wave eigenmodes of the linear ring model and the WKB analysis of Model~1 (see Section~\ref{sec:linearringandWKB}). The horizontal solid black line references zero frequency. Horizontal solid blue lines show the range in radius where the amplitude of each frequency eigenmode exceeds 10\% of its maximum value. The solid red curve corresponds to the total circular frequency curve of the linear model, $\Omega$. Dashed red illustrates the $m = 1$ vertical resonances, $\omega = \Omega \pm \nu_x$. Shaded red indicates the ``forbidden" region between the two resonances, as predicted by the WKB approximation. Evidently, the linear eigenmodes lie outside of the forbidden region and tend to lie along the two vertical resonances.
\label{fig:eigenvectors}}
\end{figure}

Following \citet{sparke1984} and \citet{sparke1988}, we divide the disc into ${\cal N}$ concentric rings.  Equations~(\ref{eq:dispersion1}) and (\ref{eq:dispersion2}) together constitute an eigenvalue equation for the $2{\cal N}$-dimensional vector $\left ({\cal Z},\,{\cal V}\right )^{T}$.  The eigenvalues $\omega_i$~ $(i=1,2{\cal N})$ are all real.  We find that each eigenmode departs significantly from zero over a relatively narrow range in $R$.  This point is illustrated in Fig.~\ref{fig:eigenvectors} for the case $m=1$ where we plot a horizontal line segment across the range in $R$ where the amplitude of the eigenmode exceeds 10\% of its maximum value. The parameters for the ${\cal N}$-ring disc and the external potential of the bulge and halo are chosen to match our simulated Model~1.

Evidently, there are two continuous branches of eigenmodes.  The
patterns of modes along the upper-$\omega$ branch rotate in the same sense
as the disc while those modes along the lower branch are
counter-rotating or rotating at very small (nearly zero) positive frequency.  In general, a disturbance localized in $R$ will
involve a superposition of modes from the two branches.  The part of
the wave associated with modes from the upper branch, where $\omega$
is a decreasing function of $R$, will shear into a trailing spiral
while the part associated with the lower branch will shear into a
leading spiral.  Since the magnitude of the differential rotation rate $|d\omega/dR|$ is
smaller along the lower branch, phase mixing occurs more slowly for
this part of the wave.  Thus, over time we expect more pronounced leading
waves of vertical oscillation, which is indeed what we find in the
simulations (cf. Figs.~\ref{fig:zmap_vmap_bar_nobar} and
\ref{fig:Zm_maps}).

For wavelengths that are small as compared with the scale length of
the disc we may use a WKB approximation.  The wavefronts are assumed to 
be perpendicular to the radial direction and the dispersion relation
becomes \citep{hunter1969, sparke1988, nelson1995}

\begin{equation}\label{eq:WKB}
\left (\omega - m\Omega(R)\right )^2 - 4\pi^2 G \Sigma(R)/\lambda -
\nu_h^2 - \nu_b^2 = 0~,
\end{equation}

\noindent where $\lambda$ is the wavelength of the perturbation.
Corotation occurs when $\omega = \Omega$, while vertical resonances
(the analogs of Linblad resonances) occur when $\omega = \Omega \pm
\nu_x$ where $\nu_x = \sqrt{\nu_b^2 + \nu_h^2}$ is the vertical oscillation
frequency associated with the halo and bulge (i.e., the fixed spheroidal
components).  From equation~(\ref{eq:WKB}), we see that waves are
excluded from the region in frequency space between these two
resonances, the so-called ``forbidden" region \citep[][Section 6.6.1]{nelson1995, binney2008}.

In addition to the results of the eigenmode analysis \citep{sparke1988} in Fig.~\ref{fig:eigenvectors}, we also show the corotation and vertical resonance curves as calculated in the WKB approximation. Indeed, the linear eigenmodes tend to lie along the two vertical resonances.

\section{Corrugations in External Galaxies}
\label{sec:externalwaves}

If warps and corrugations in Milky Way-like disc galaxies are easily excited and long
lived then they should be observable in similar external galaxies. As shown in
this work, and demonstrated in others \citep[for example,
see][]{gomez2017}, spatial corrugations in the vertical direction are
coupled to corrugations in stellar vertical velocity. Furthermore,
\citet{gomez2017} showed that vertical perturbations in simulated disc galaxies
manifest in both the gaseous and stellar components. Thus, velocity
corrugations should be observable in the line-of-sight kinematics of
nearly face-on external disc galaxies. Integral field spectroscopy
surveys such as Calar Alto Legacy Integral Field Area
\citep[CALIFA;][ but also see
\citealt{falcon-barroso2017}]{sanchez2012}, Mapping Nearby Galaxies at
APO \citep[MaNGA;][]{bundy2015}, and DiskMass \citep{bershady2010} hold great promise for detecting bending waves in external
galaxies since they gather spatially resolved information on both the stellar and gaseous
components.  It is most likely easier to detect any corrugations in
the gaseous disc rather than the stellar component due to larger
velocity dispersions in the latter. In this Section we outline a
number of requirements needed to observe extragalactic
bending waves and corrugations with respect to our simulations.

\begin{figure}
\includegraphics[width=\columnwidth]{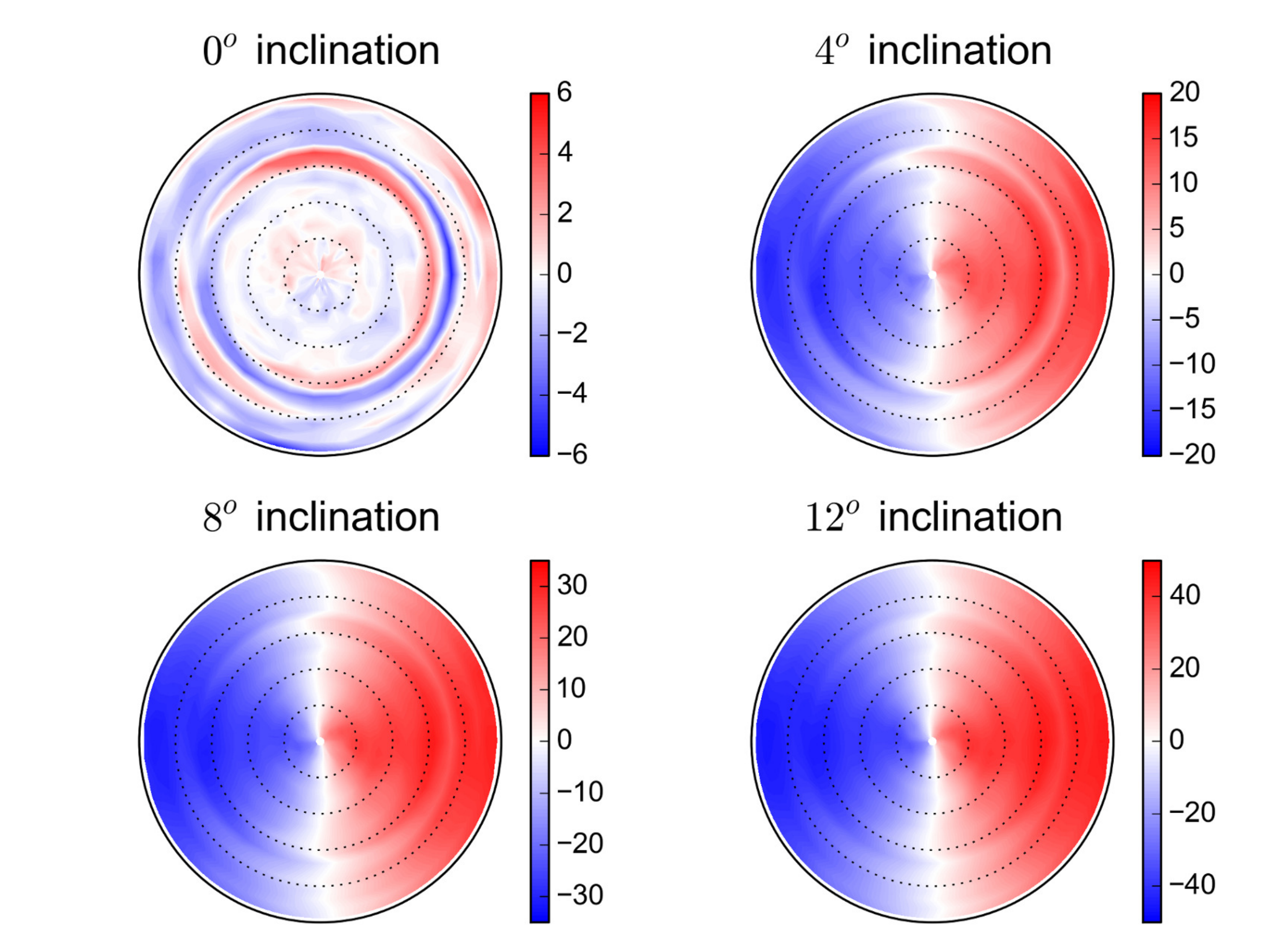}
\caption{Projected line-of-sight velocity field of Model~1 at $t \sim 3.5 \, {\rm Gyr}$ for four inclinations, as indicated in the figure, rotated about the horizontal axis. Dotted concentric circles indicate increments of 5 kpc in projected radius. The rotation of the disc is counter-clockwise in the face-on reference frame.   The colour scale is in units of ${\rm km} \, {\rm s}^{-1}$, and differs in each panel to highlight the change in amplitude. Contributions from both radial and azimuthal flows become greater along the line-of-sight as the inclination increases. Despite this, hints of corrugated patterns are still apparent (for example, for $R$ between $\sim 15 - 20 \, {\rm kpc}$), even at the largest inclination shown in the figure.
  \label{fig:inclination_observations}}
\end{figure}

In Fig.~\ref{fig:inclination_observations} we show line-of-sight
velocity maps for our Model~1 as viewed face-on and at inclinations
$i=4^\circ,\, 8^\circ,\,$ and $12^\circ$.  Contributions to the line-of-sight velocity from both radial and azimuthal flows increase with
increasing inclination.  Nevertheless, hints of corrugation are still
visible even at $i=12^\circ$.  In principle, one might attempt to
incorporate vertical motions into a model for two-dimensional
velocity maps of spiral galaxies, as is done for non-circular 
flows in the presence of a bar \citep{spekkens2007,sellwood2010,sellwood2015}.

\begin{figure}
\includegraphics[width=\columnwidth]{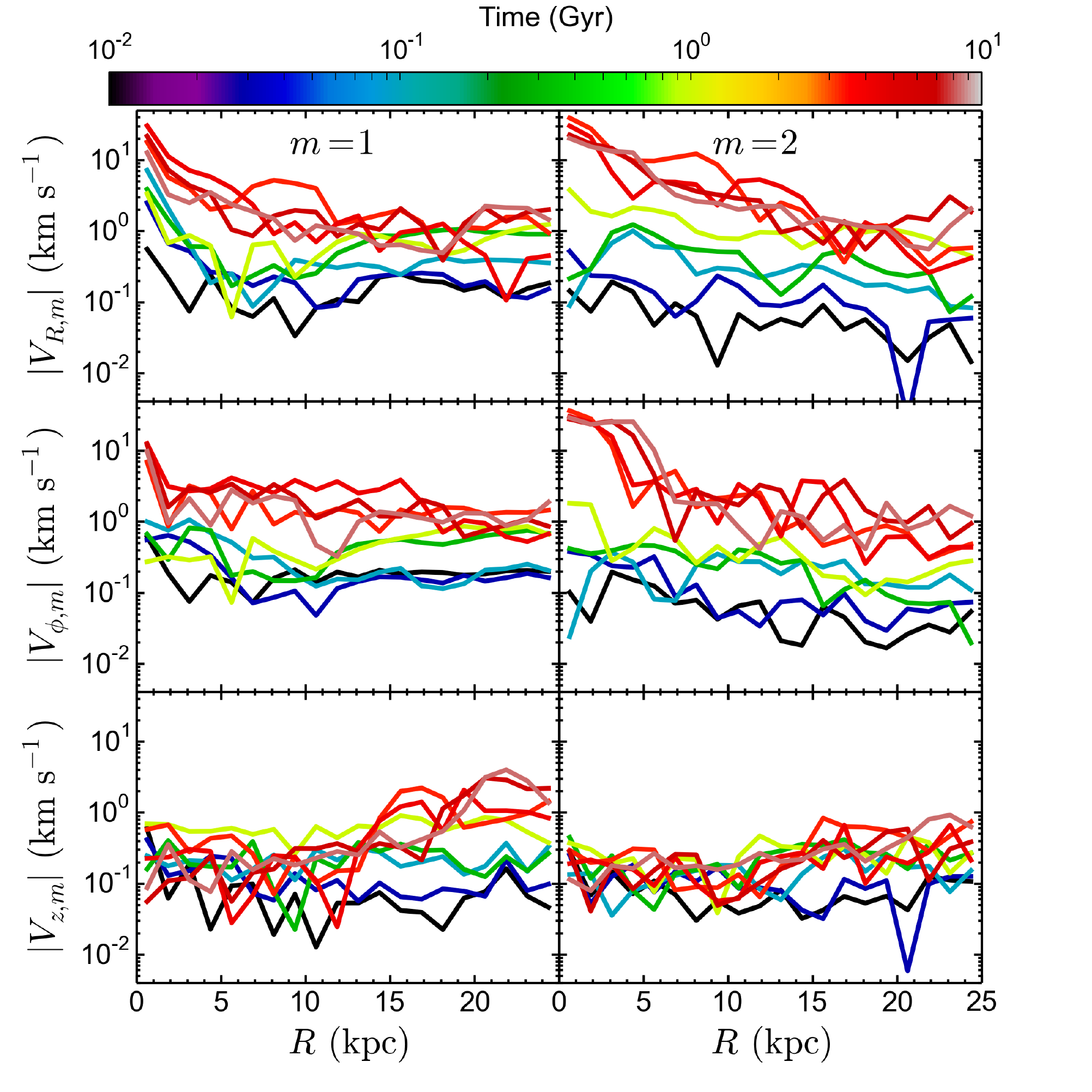}
\caption{Time evolution of the $m = 1$ and $m = 2$ (left and right columns, respectively) Fourier coefficient magnitude as function of radius for cylindrical radial, azimuthal, and vertical velocities (top, middle, and bottom rows, respectively), for Model~1. Colour indicates the epoch of each magnitude's profile. The large in-plane $m = 2$ amplitudes in the central disc are due to the bar, while the large $m = 1$ amplitude of vertical velocity in the outer disc at later times is a manifestation of the warp and other vertical corrugation patterns. As the disc is projected onto the sky and inclined the $m = 1, \, 2,$ and higher order terms (not shown) mix into the line-of-sight velocity. In the text we discuss at what inclinations the $m=1$ vertical motions dominate the mixed non-circular in-plane motions in the line-of-sight velocity field.
\label{fig:VtVRpowerspectra}}
\end{figure}

We may thus ask, at what inclination do the motions in the vertical
direction dominate the in-plane flows along the line-of-sight?  That
is, how face-on does a galaxy need to be?  In
Fig.~\ref{fig:VtVRpowerspectra} we consider the radial, azimuthal, and
vertical bulk motions for Model~1.  In particular, we show a time
sequence of $m=1$ and $m=2$ Fourier mode amplitudes as a function of
radius (i.e. equations~\ref{eq:displacement} and
\ref{eq:bendmodefouriercoef}, but for velocity rather than vertical
displacement).  In the inner disc, the largest amplitudes correspond to the $m=1$ and $m=2$ radial and azimuthal velocities, and are due to the bar.  On the
other hand, large $m=1$ vertical velocity amplitudes in the outer disc, especially at
late times, are due to the warp and associated corrugations.  

For an inclined disc, the $m=1$ and $m=2$ contributions, as well as
higher $m$ terms from lower amplitude waves, mix into the
line-of-sight velocity field. After accounting for the systematic velocity of a galaxy, the line-of-sight velocity can be decomposed as

\begin{equation}
v_{\rm los} = (v_R {\rm sin} \, \theta + v_{\phi} {\rm cos} \, \theta ) \, {\rm sin} \, i + v_z {\rm cos} \, i~,
\end{equation}

\noindent
where $v_R$, $v_{\phi}$, and $v_z$ are Fourier expansions of each respective velocity component, $\theta$ is the position angle of the disc, and $i$ is the inclination. In Fig.~\ref{fig:VtVRpowerspectra}, non-circular radial and azimuthal flows roughly equal that of the $m=1$ vertical component in the outer disc. Thus, upon averaging over the position angle term on the interval $[0,\pi]$, bulk $m=1$ vertical motions in the outer disc will be comparable to the integrated $m=1$ and $m=2$ in-plane contributions along the line-of-sight for an inclination of $\sim 38 \degr$. For inclinations of $\sim 20 \degr$ and  $10 \degr$, the average vertical contribution to the line-of-sight velocity field would be roughly double and four times that, respectively, of the non-circular in-plane flows.

In addition to an inclination constraint, a very large field of view
is needed in order to resolve the velocity corrugations that we find
in the outer disc. The $m=1$ bending
waves for Model~1 (Fig.~\ref{fig:VtVRpowerspectra}) dominate in the region $R \sim
20 - 25 \, {\rm kpc}$, which corresponds to $\sim 6 - 8$ disc scale
lengths, or $\sim 4 - 5$ effective radii (assuming an exponential disc
and $1 R_e \simeq 1.7 R_d$).  Furthermore, the spatial resolution
required to resolve corrugations similar to those presented in this
work should be at least $\sim 2 - 3 \, {\rm kpc}$.

The presence of strong spiral arms may complicate the interpretation
of any observed ``vertical motions''.  Since stars on the side of a
face-on disc furthest from an observer are obscured by dust,
compression and rarefaction of the disc perpendicular to the disc
plane will appear as bulk vertical motions.  Spiral arms can generate such breathing wave motions in the disc, and even arms with modest density contrasts can produce bulk motions comparable to what we observe in this work \citep{debattista2014,faure2014,monari2016a,monari2016b}. It could be that the corrugations observed by \citet{sanchez-gil2015} using long slit spectroscopy are associated with breathing mode perturbations from the spiral structure. However, the vertical motions will trace the spiral arms and follow a distinct compression and rarefaction pattern on either side of the arms. The bending waves we observe in our simulations are tightly wound and leading, and therefore should be distinguishable from bulk motions induced by (trailing) spiral arms.

\section{Discussion}
\label{sec:discussion}

A satellite or dark matter subhalo that passes through a stellar disc will set up disturbances in the disc, which include vertical bending and breathing waves. The simulations presented in this paper demonstrate that bending waves also arise in an isolated Milky Way-like stellar disc without provocation from an outside agent such as a passing satellite.

At intermediate radii ($R \sim 8-15 \, {\rm kpc}$ for a Milky Way-like galaxy), these waves develop into tightly wound leading spirals with an amplitude and radial wavelength that is similar to those recently discovered in the Milky Way \citep{xu2015}. In contrast to the hypothesis of \citet{xu2015} that the corrugated structures are associated with spiral arms, we find that they are independent, both in morphology and pattern speed.

The bending waves in our simulations persist for many billions of years.  Over time, bending wave power migrates outward.  The warp begins at $R \sim 10 - 15 \, {\rm kpc}$ with an amplitude of $\sim 200 \, {\rm pc}$ that increases to $\sim 300 - 500 \, {\rm pc}$ closer to the disc's edge. These amplitudes are in agreement with recent measurements of the stellar warp in the Milky Way \citep[see for example][ and references therein]{reyle2009}. The outward migration of power is more dramatic for Model~1 where the bar and/or more massive disc plays a dominant role in suppressing bending waves, especially at intermediate radii. Since our Galaxy has a central bar, the implication is that bending wave corrugations are efficiently damped in the inner disc, and should only be detected at Galactocentric radii greater than that of the Sun.

What then is the origin of the bending waves in our simulated galaxies?  A likely culprit is the random noise of the halo and bulge particle
distributions.  To test this hypothesis, we re-simulated Model~1 with a
fixed analytic potential for the bulge and halo and found that the
vertical perturbations were negligible even though the disc developed
spiral structure and a bar. We note bar formation was delayed as compared with the live halo case \citep[see][]{sellwood2016}.  As a further check, we evolved the live disc
in a fixed potential that was generated by halo and bulge particles
frozen in their initial positions.  In this case, the disc developed a
strong corrugation pattern of bending waves at intermediate radii and
a large warp near the edge of the disc.  Evidently, large-scale vertical waves can be excited by particle noise in the halo but not by structures in the disc itself, such as a bar or spiral arms.

Once formed, the evolution of bending waves is remarkably well
described by linear perturbation theory, especially during the early
stages of the simulations.  The linear eigenmodes of a razor thin, dynamically
cold disc form a continuum of bending waves, rather than discrete modes, with rotational frequencies that
lie along the vertical resonance curves at $\omega = \Omega \pm \nu_x$
\citep{hunter1969, sparke1988}.  Of course, our $N$-body discs have a
non-zero vertical velocity dispersion and thickness. Also, the subsequent evolution of our live models obviously differs from that in the linear regime, where dynamical friction from the halo tends to efficiently damp bending waves on very short time-scales \citep{nelson1995}. Furthermore, as the disc evolves, in-plane and vertical (time-dependent) inhomogeneities and asymmetries emerge, driven by the bar, spiral structure, mass redistribution within the disc, and resolution effects, which the linear theory and WKB analysis does not incorporate. Nevertheless,
power in the $R-\omega$ plane concentrates along the two resonances
and lies, for the most part, outside the region between them (i.e.,
the forbidden region as predicted by WKB theory). 

Since both linear theory and the WKB analysis include self-gravity of the perturbations, we conclude that the phenomena observed in our discs are true bending waves. This contrasts with kinematical phase-wrapping explanations of the observed local bending and breathing modes in the Solar Neighbourhood as suggested by \citet{delavega2015}. However, \citet{delavega2015} studied bending and breathing waves that arose from satellite interactions with the disc and close flybys of a fairly massive Sagittarius dwarf model. Admittedly, we do not consider direct interactions between the disc and external perturbing agents in this paper. One could imagine a two-stage scenario in which the waves present at the early stages of a satellite encounter behave at least somewhat kinematically. Later on, when these perturbations have phase-mixed and sheared into extended arcs they can be described by a ring model, in which self-gravity is a key ingredient.

We also conclude that our bending waves are not strongly Landau-damped, as one might have anticipated from the analysis of vertical oscillations in one-dimensional systems \citep{mathur1990, weinberg1991, louis1992, widrow2015}.  Likewise, the waves are not strongly damped by dynamical friction due to the halo.  If anything, a live halo (and bulge) appear to give rise to more vigorous perturbations perpendicular to the disc as predicted by \citet{bertin1980} and in discord with the analysis of \citet{nelson1995}.

Our simulations are consistent with the picture presented in
\citet{binney1998}.  Recall that the authors of that paper simulated a
disc-halo system where the disc was represented by rigid concentric
rings while the halo constituted standard $N$-body particles.
They showed that if the disc was initialized in the configuration of a
discrete warp mode
\citep[e.g., the modified-tilt mode of][]{sparke1988} the warp rapidly winds up while retaining its warp
energy.  They conclude that the true modes of the disc-halo system are
different from those of the disc-static halo system, essentially
because the halo responds to the warp and thus changes the potential
in which the disc evolves.  The spectral analysis of our simulations
suggests that while the modes of the disc-halo system might differ
quantitatively from those of the disk-static halo system, the general
structure of the modes, at least those of the continuum, are
qualitatively similar.  The key point is that there exists a low-frequency branch where the differential precession is small at larger radii. Thus, the winding problem is alleviated and bending waves persist for many dynamical times.

One might expect, then, that waves at smaller radii would be sheared out via differential rotation. The survival of these waves in our simulations, especially those along the low-frequency branch, is likely attributed to the strong self-gravity of the inner disc. Moreover, random in-plane motions act to further stiffen the inner disc and resist bending \citep{debattista1999}. A straight LON and constant pattern speed in the inner disc indicates that the waves rotate coherently and cohesively, connecting the global structure of low-frequency waves between the inner and outer disc. The waves in the inner disc of our simulations differ from that predicted by the linear ring model since neighbouring rings cannot be considered to oscillate independently and epicyclic motions are not taken into account. This point implies at least some degree of non-linearity present in our simulations regarding the bending waves -- the waves we see may very well be akin to the true modes of the disc hinted at by \citet{binney1998}.

\section{Conclusions}
\label{sec:conclusions}

Our main conclusion is that bending waves should be generic,
long-lived features of Milky Way-like disc galaxies irrespective of whether the disc
has been perturbed by a passing satellite or dark matter subhalo.
This conclusion is based on the results from N-body simulations of
isolated Milky Way-like galaxies.  In particular, we focus on two
models: one with a maximal bar-forming disc and the other with a
submaximal disc which only forms flocculent spiral structure.  In
both models, bending waves develop across the entire disc within the
first billion years of the simulation. At intermediate radii the waves manifest as leading, tightly wound corrugations, which extend over a large range in azimuth and match smoothly on to the warp that develops at the edge of the disc. 

A major goal of this paper is to develop tools for the study of
bending waves in simulations and observations.  Our approach is based
on the spectral analysis of surface density maps for simulated disc
galaxies -- a tool that has proved indispensable for understanding the
dynamics of bars and spiral structure.  In principle, the complete
dynamical state of a stellar disc is encapsulated in the DF, a
complicated function of the six phase space coordinates and time.  To
simplify the analysis, one can derive a surface density map,
essentially by integrating the DF over ${\bf v}$ and $z$.  The
standard procedure is to then divide the disc into radial bins and
write the surface density in each bin as a series of functions with
$m$-fold azimuthal symmetry.  The coefficients provide the relative
strength of the different angular modes as a function of $R$ and $t$.
By Fourier transforming in time, we can also obtain a power spectrum
in $R$ and $\omega$.

For this work, we begin with the $z$ and $v_z$ moments of the DF
across the disc, which provide a measure of bending waves in the disc.
As with surface density maps, we can decompose maps of $\langle
z\rangle$ and $\langle v_z \rangle$ as Fourier series in $\phi$.  We
focus on the $m=1$ waves, which is the dominant term not only for the
warp but also for the bending waves that occur at intermediate radii.
Furthermore, the $m=1$ waves provide a direct link to tilted
ring models for warped galaxies and the eigenfunction calculations of
\citet{hunter1969} and \citet{sparke1988}. From Fig.~\ref{fig:PS_evolution},  we conclude that at a given $R$, the amplitude of the $m=1$ bending waves
grows with time until reaching some maximal value, after which it
remains roughly constant.  The maximum amplitude is an increasing
function in $R$ as is the time at which it is reached.  The upshot is that over time, the strength of the bending waves near the edge of the disc increases relative to that of waves in the inner part of the disc.

The long-lived nature of bending waves is consistent with the
conclusions of \citet{binney1998}, who showed that a live halo
responds to vertical displacements on a relatively short time-scale.
In essence, bending waves are a phenomena of the disc-halo system.  In the absence of this coupling between the disc and live halo, dynamical friction would efficiently damp the bending waves, as was suggested by \citet{nelson1995}.

The connection between linear theory, in the form of either
eigenfunction or WKB analyses, and the simulation results is most
clearly seen in Figs.~\ref{fig:powertimeseries} and \ref{fig:eigenvectors}. The agreement is impressive, as power in the $m=1$ waves tends to reside
just outside the WKB-forbidden region between the two vertical
resonance curves.  At early times, the power is distributed
approximately uniformly throughout the disc while at late times power is concentrated near the outer edge of the disc, lending credence to an inside-out formation scenario of the warp.  We emphasize that central to the linear theory calculations is a gravitational restoring force that acts on perturbations in the disc.  Hence, the agreement between simulations and linear theory bolsters our claim that we are observing (at least in the simulations) bona fide gravity waves.

It was a surprise (at least to us) that bending waves arose in a disc embedded in a smooth dark matter halo.  Evidently, particle noise in the halo is enough to provoke vertical waves in a stellar disc.  Of course, dark matter haloes in $\Lambda$CDM cosmologies are predicted to host a system of satellite galaxies and dark matter subhaloes (for example, see the early works of \citealt{klypin1999} and \citealt{moore1999}). In a realistic galaxy, a small fraction of these halo objects will pass through the disc, an
interaction that is characterized by a series of phases.  In the
first, a local region of the disc is perturbed by the passing
object.  The response of individual stars will depend on how well
their vertical epicyclic motions match the time-dependence of the
gravitational field \citep{sellwood1998,widrow2014}.
The localized perturbation is then sheared out due to differential
rotation and phase mixing \citep{widrow2014,delavega2015} but soon begins to behave as a tightly wound bending wave with gravity providing the restoring force.  In essence, this is the phase investigated here.  Eventually, the wave energy is dissipated through further phase mixing and Landau damping. Consequently, the disc then becomes a little thicker and kinematically hotter. Using the shot noise in our simulations as a proxy for asymmetries in haloes, it is tangible to imagine bending waves in Milky Way-like (stellar) discs to be continually excited.

One of the goals of the \textit{Gaia} observing mission \citep[][but see \citealt{gaiacollaboration2016b} for an overview of DR1]{perryman2001,gaiacollaboration2016a} is to provide a map of the mean velocity field over a substantial fraction of the Galactic disc.  The picture presented here suggests that this map will reveal a mix of localized flows and larger scale motions that connect waves at intermediate radii with the outer warp (see, for example, \citealt{abedi2014} regarding \textit{Gaia} and the Galactic warp). In principle, the former will tell us something about perturbers roaming the halo while the latter may say more about the structure of the disc. We predict that less dispersive low-frequency $m=1$ bending waves, which manifest as tightly wound and leading radially dependent vertical asymmetries in position and velocity, should be prevalent at radii just beyond the Sun. Although similar features have already been observed in the Milky Way's density distribution \citep{xu2015}, the key will be testing the true wave nature of these patterns using proper motion and radial velocity measurements from the next \textit{Gaia} data release. In theory, the analysis tools based on moments in $z$ and $v_z$, and azimuthal Fourier modes, can be applied to \textit{Gaia} data, though there, one must contend with selection effects, observational uncertainties, and obscuration due to dust.

One possible complication, not discussed here, is that different components of the disc may respond differently to a passing satellite.  \citet{bovy2013} have suggested that to do an Oort type analysis (that is, to infer the surface density, vertical force, and density of dark matter in the disc) one should divide disc stars into bins based on [$\alpha$/Fe] and [Fe/H] -- the idea being that the stars in each bin can be treated as a distinct isothermal tracer of the gravitational potential.  On the other hand, if the disc is in a continual state of disequilibrium, and if the manifestation of disequilibrium varies from one population to the next, then population-dependent systematic errors might creep into this sort of analysis \citep{banik2017}.  Put another way, discrepancies in the inferred potential might signal that the Galaxy is in a perturbed state.

In summary, \textit{Gaia} may well find complicated patterns in the bulk motions of disc stars.  The challenge will then be to characterize these motions, perhaps with the spectral methods described here, and disentangle the effects of external perturbers and internal dynamics.

\section*{Acknowledgements}

The authors would like to thank the anonymous referee for helpful comments that improved the quality of the manuscript. The authors are also grateful to Heidi Newberg, Kristine Spekkens, and Sebasti{\'a}n S{\'a}nchez for useful conversations, and Jacob Bauer for his assistance with the simulations. We also acknowledge the use of computational resources at the Centre for Advanced Computing. This work was supported by the Natural Sciences and Engineering Research Council of Canada through the Postgraduate Scholarship Program (MHC) and Discovery Grant Program (LMW).

%%%%%%%%%%%%%%%%%%%%%%%%%%%%%%%%%%%%%%%%%%%%%%%%%%

%%%%%%%%%%%%%%%%%%%% REFERENCES %%%%%%%%%%%%%%%%%%

% The best way to enter references is to use BibTeX:

\bibliographystyle{mnras}
\bibliography{bibliography}

% Don't change these lines
\bsp	% typesetting comment
\label{lastpage}
\end{document}